\documentclass[twocolumn]{aastex61}
\usepackage{amstext}
\usepackage{amsmath}
\usepackage{amssymb}
\usepackage{apjfonts}

\shorttitle{Common Envelope Wind Tunnel}
\shortauthors{MacLeod et al.}

\begin{document}

\newcommand{\beq}{\begin{equation}}
\newcommand{\eeq}{\end{equation}}
\newcommand{\beqar}{\begin{eqnarray}}
\newcommand{\eeqar}{\end{eqnarray}}

\newcommand{\Ra}{R_{\rm a}}
\newcommand{\Ma}{M_{\rm a}}
\newcommand{\vinf}{v_\infty}
\newcommand{\rhoinf}{\rho_\infty}
\newcommand{\mach}{\mathcal M}
\newcommand{\mhl}{\dot M_{\rm HL}}
\newcommand{\lhl}{\dot L_{\rm HL}}
\newcommand{\Hrho}{H_\rho}
\newcommand{\Rs}{R_{\rm s}}
\newcommand{\erho}{\epsilon_\rho}
\newcommand{\cs}{c_{\rm s,\infty}}
\newcommand{\ehl}{\dot E_{\rm HL}}
\newcommand{\vk}{v_{\rm k}}
\newcommand{\fk}{f_{\rm k}}
\newcommand{\gs}{\Gamma_{\rm s}}
\newcommand{\g}{\gamma}

\title{Common Envelope Wind Tunnel: Coefficients of Drag and Accretion in a Simplified Context for Studying Flows Around Objects Embedded Within Stellar Envelopes}

\correspondingauthor{Morgan MacLeod}
\email{morganmacleod@ias.edu}

\author{Morgan MacLeod}
\altaffiliation{NASA Einstein Fellow}
\affiliation{School of Natural Sciences, Institute for Advanced Study, 1 Einstein Drive, Princeton, New Jersey 08540, USA}

\author{Andrea Antoni}
\affiliation{Department of Astronomy \& Astrophysics, University of California, Santa Cruz, CA 95064, USA}

\author{Ariadna Murguia-Berthier}
\affiliation{Department of Astronomy \& Astrophysics, University of California, Santa Cruz, CA 95064, USA}

\author{Phillip Macias}
\affiliation{Department of Astronomy \& Astrophysics, University of California, Santa Cruz, CA 95064, USA}

\author{Enrico Ramirez-Ruiz}
\affiliation{Department of Astronomy \& Astrophysics, University of California, Santa Cruz, CA 95064, USA}

\begin{abstract} 
This paper examines the properties of flows around objects embedded within common envelopes in the simplified context of a ``wind tunnel.'' We establish characteristic relationships between key common envelope flow parameters like the Mach number and density scale height. Our wind tunnel is a three-dimensional, cartesian geometry hydrodynamic simulation setup that includes the gravity of the primary and secondary stars and allows us to study the coefficients of drag and accretion experienced by the embedded object. Accretion and drag lead to a transformation of an embedded object and its orbit during a common envelope phase. We present two suites of simulations spanning a range of density gradients and Mach numbers -- relevant for flow near the limb of a stellar envelope to the deep interior. In one suite, we adopt an ideal gas adiabatic exponent of $\gamma=5/3$, in the other, $\gamma=4/3$. We find that coefficients of drag rise in flows with steeper density gradients and that coefficients of drag and accretion are consistently higher in the more compressible, $\gamma=4/3$ flow. We illustrate the impact of these newly derived coefficients by integrating the inspiral of a secondary object through the envelopes of $3M_\odot$ ($\gamma\approx5/3$) and $80M_\odot$ ($\gamma\approx4/3$) giants. In these examples, we find a relatively rapid initial inspiral because, near the stellar limb, dynamical friction drag is generated mainly from dense gas focussed from deeper within the primary-star's envelope. This rapid initial inspiral timescale carries potential implications for the timescale of transients from early common envelope interaction. 
\end{abstract}

\section{Introduction}

Common envelope episodes occur in binary star systems when one star engulfs its companion \citep{1976IAUS...73...75P}. These episodes, during which drag on the surrounding gas transforms and tightens the binary orbit, are thought to be critical in the formation of compact binaries. In particular, binaries that are able to merge under the influence of gravitational radiation (like the recently discovered merging pairs in the LIGO experiment) must be brought to separations smaller than the size of the stars that created them \citep[e.g.][]{2007PhR...442...75K,2016ApJ...818L..22A}. This can only occur through a phase of orbital transformation like a common envelope episode. The details of this common envelope interaction phase therefore determine the exact nature of the resultant binary \citep{1993PASP..105.1373I,2000ARA&A..38..113T,2010NewAR..54...65T,2013A&ARv..21...59I,2014LRR....17....3P}. 

A typical common envelope event is thought to pass through several phases \citep{2001ASPC..229..239P,2010NewAR..54...65T}. Over stellar evolution ($\gtrsim 10^5$~year) timescales, a star in a pair evolves from its main-sequence size onto the giant branch, growing in radius significantly. Along the way it may become so large that it starts to impinge on the orbit of its companion star, when the orbital separation is similar to the stellar radius, $a\sim R_*$. This growth initiates interaction between the pair, which exponentiates through  exchange or loss of mass (and angular momentum) from the system \citep{2016MNRAS.455.4351P,2016MNRAS.461.2527P}, or by tidal instability (the Darwin instability, which is relevant mainly in pairs of unequal mass) \citep[e.g.][]{2016arXiv160501493M}. Thus the onset of the interaction occurs over a timescale regulated either by mass loss or by tidal dissipation, perhaps lasting hundreds to thousands of orbital periods in either case \citep[e.g.][]{2011A&A...528A.114T,2014ApJ...786...39N,2016arXiv160501493M}. Both of these processes desynchronize the orbit of the secondary star from that of the primary's envelope.  

Eventually, one object is engulfed within the envelope of the other and the common envelope phase begins. Supersonic relative motion between the engulfed object and the envelope gas leads to gravitational focussing and the buildup of a dense wake behind the embedded object, which exerts a gravitational drag on the orbital motion \citep{2013A&ARv..21...59I}. The result of this `dynamical friction' drag is a rapid inspiral through the increasingly dense stellar envelope. This phase, sometimes called the dynamical plunge, has two possible conclusions. In some cases, the pair of stars merge. In others, the two stellar cores both heat a fraction of envelope material, resulting in subsonic relative motion between the embedded objects and the gas. With this transition to subsonic relative velocity, drag forces drop off dramatically \citep[e.g.][]{1999ApJ...513..252O} and the new binary's orbit stabilizes \citep[e.g.][]{2012ApJ...746...74R,2012ApJ...744...52P,2016ApJ...816L...9O,2017MNRAS.464.4028I}.

Despite significant recent effort and progress \citep[e.g.][]{2008ApJ...672L..41R,2012ApJ...746...74R,2012ApJ...744...52P, 2014ApJ...786...39N,2015MNRAS.450L..39N,2016MNRAS.460.3992N,2016MNRAS.462..362I,2016ApJ...816L...9O,2016MNRAS.462L.121O,2016arXiv161200008O,2016MNRAS.455.3511S,2016MNRAS.458..832S,2017MNRAS.464.4028I}, global simulations of common envelope remain challenging to perform with many potentially important processes and timescales at play.  A particular concern is the resolution requirement of simulating the full spatial extent of the binary for many orbital timescales implies that the simulations are either very computationally expensive, allowing a small number of calculations to be performed \citep[e.g.][]{2017MNRAS.464.4028I}, or they are performed with extremely low numerical resolution but can span some parameter space \citep[e.g.][]{2016MNRAS.462..362I}. 
As a result of these numerical limitations and physical complexity, fully interpreting and learning from the results of these simulations has proven to be challenging.

In this paper, we adopt the complementary approach of studying a well-defined, but idealized scenario related to common envelope encounters in detail. We model flow past a gravitating object like one embedded in a common envelope phase in the context of  a ``wind tunnel" numerical setup. 
By restricting the scope of the problem from the global scenario, this idealized approach allows us to examine the importance of individual physical processes separately from the full, complex system. These results, in turn, can prove valuable in interpreting the findings of global calculations. 
Our work builds on a long history of study of supersonic flows past gravitating objects, starting with \citet{1939PCPS...35..405H} and \citet{1944MNRAS.104..273B,1952MNRAS.112..195B}.
The analytic scalings of these flows have also informed a large portion of our understanding of the hydrodynamics of common envelope interactions \citep[see, for example][]{1988ApJ...329..764L,1993PASP..105.1373I,2013A&ARv..21...59I}.

Numerical studies have augmented this analytic understanding beginning with pioneering simulations by \citet{1971MNRAS.154..141H}. Later, work by \citet{1985MNRAS.217..367S} was the first to solve for the properties of this flow with a finite volume computational method. This work also took the important step of calculating coefficients of dynamical friction drag due to the gravitational interaction of the object with its wake. Subsequent work studied flows in inhomogeneous media, making the results more directly applicable to understanding flow around objects embedded in the common envelope \citep{1986MNRAS.218..593L,1986MNRAS.221..445S,1986MNRAS.222..235L,1987ApJ...315..536F,1988ApJ...335..862F,1989ApJ...339..297T,2000ApJ...532..540A}.

Numerical advances facilitated what remains a benchmark series of simulations of Hoyle-Lyttleton flow by Ruffert \citep{1994ApJ...427..342R,1994ApJ...427..351R,1994A&AS..106..505R,1995A&AS..113..133R,1996A&A...311..817R,1997A&A...317..793R,1999A&A...346..861R}. This work is notable for its relatively high numerical resolution, and for being the first broadly successful attempt to span a wide parameter space of flow mach numbers, adiabatic exponents, and object sizes. 
\citet{2009ApJ...700...95B} and \citet{2012ApJ...752...30B} show that with high resolution and modern numerics, three-dimensional flows in homogenous media are stable and reach a steady state with accretion rates on the order of the \citet{1939PCPS...35..405H} estimate.\footnote{See section 3.1 of \citet{2015ApJ...803...41M} for a more detailed discussion of the numerical assumptions and results of this recent work.} While accretion and flow morphology are the focus of much of the above work, dynamical friction drag forces have also been a focus of recent numerical studies, some of which have conditions that are  particularly relevant to common envelope flow \citep[in particular,][]{1999ApJ...522L..35S,2009ApJ...703.1278K,2012ApJ...745..135S,2016A&A...589A..10T}.

This paper adds to this history of numerical study of Hoyle--Lyttleton and related flows and their application to the common envelope phase of binary interaction. To do so, we expand on an idealized formalism for studying the dynamical inspiral phase of common envelope introduced in \citet{2015ApJ...803...41M} and model flow past an object embedded in a numerical ``wind tunnel.''  
To determine the conditions of the wind, we consider stellar structures (and gas adiabatic exponents) relevant to two key regimes of common envelope encounters. In the convective envelopes of low-mass stars, gas pressure dominates and a $\gamma = 5/3$ equation of state describes the gas well. In higher mass stars, radiation pressure is quite important and the gas response to compression is closer to $\gamma=4/3$. This paper examines both regimes (for encounters with a 1:10 mass ratio) and compares flow properties in each. 

The remaining sections of this paper are organized as follows. In Section \ref{sec:flowconditions} we introduce key descriptive parameters for common envelope flows and derive relationships between them. These flow properties inform the setup of our numerical experiments -- which we call the Common Envelope Wind Tunnel. Section \ref{sec:numericalmethods} describes our numerical method. Section \ref{sec:numericalresults} examines the results of a set of numerical experiments comparing flows with $\gamma = 5/3$ to those with $\gamma = 4/3$.  Section \ref{sec:implications} illustrates the implications that these idealized results have for our understanding of the nature of typical common envelope inspirals. 
In Section \ref{sec:conclusion} we conclude.

\section{Flow Conditions During Common Envelope Inspiral}\label{sec:flowconditions}

\subsection {Characteristic Scales}\label{sec:scales}
Let us imagine the interaction between a giant-star primary with total mass $M_1$ and radius $R_1$ with a secondary object of mass $M_2$ and radius $R_2$, which will become embedded within the primary. The separation between these two objects is $a$, and during the interaction, $a<R_1$. We define the mass ratio of the system as $q=M_2/M_1$.
The characteristic orbital velocity is 
\beq\label{vkdef}
\vk = \left( G M \over a  \right)^{1/2}
\eeq
where $M = M_1 + M_2$. In general, the orbital motion of $M_2$ is desynchronized from the primary's gaseous envelope and the relative velocity will be written as $\vinf = \fk \vk$, where $\fk$ is the fraction of keplerian velocity that describes the relative motion between the secondary object and the gas.

A long-standing conceptual framework for understanding flows during the dynamical plunge phase of the common envelope inspiral has been that of \citet{1939PCPS...35..405H} accretion flows \citep[e.g.][]{1978ApJ...222..269T,1979A&A....78..167M,1988ApJ...329..764L,1988ApJ...335..862F,1991ApJ...383..761K,1993PASP..105.1373I,1993ApJ...411L..33C,1995ApJ...440..270B,2013A&ARv..21...59I}. In these cases gas moves supersonically  past a gravitating object. The Mach number is
\beq\label{machdef}
\mach = {\vinf \over \cs}.
\eeq
Gravitational focussing leads gas within an impact parameter, 
\beq\label{radef}
\Ra = \frac{2 G M_2}{\vinf^2}  ,
\eeq
to be energetically bound to the accreting object of mass $M_2$. Note that the simple expression above of \citet{1939PCPS...35..405H}, and later \citet{1944MNRAS.104..273B}, ignores the gas internal energy, which is added in the \citet{1952MNRAS.112..195B} formalism \citep[see][for a review]{2004NewAR..48..843E}.

The relationship between $\Ra$ and $a$ is dictated by the relative masses in the system and fraction of Keplerian rotation,
\beq\label{raovera}
{\Ra \over a} = {2 \over \fk^2} {M_2 \over M}  = {2 \over \fk^2} {1 \over 1+ q^{-1}} .
\eeq
This ratio describes the relative size of an accretion structure to the size of the binary orbit. Note that in the simplifying case of $\fk = 1$ and $M_1 \gg M_2$, then $\Ra/a \approx 2 q$. 

A final length scale that plays a role in defining the common envelope interaction is the density scale height,
\beq\label{hrho}
H_\rho = - {\rho} \left({d\rho  \over d r} \right)^{-1}   ,
\eeq
where $d\rho/dr$ describes the density profile within the primary star's envelope \citep[e.g.][]{2015ApJ...803...41M}.  
Another important ratio describes the density gradient across the accretion radius. We define the ratio 
\beq\label{erhodef}
\erho = {\Ra \over H_\rho}
\eeq
to describe the number of density scale heights subtended by the accretion radius ($\erho \rightarrow 0$ describes a homogeneous density structure, where $\erho \rightarrow \infty$ describes a very steep density gradient). 

Taken together, the accretion radius, Mach number, and density gradient describe the hydrodynamic properties of common envelope flows, which we will focus on in this paper. 

\subsection{Polytropic Stellar Envelopes}\label{sec:polytropes}

To obtain approximate profiles of the stellar envelope structure into which the secondary star plunges in the common envelope event, we will consider polytropic envelope profiles in hydrostatic equilibrium. Here we will also make the approximation of a mass-less envelope (which implies that the bulk of the mass is concentrated in the giant-star core). This is a crude approximation of stellar structure, but one that still yields useful results (as we will show in the following sections).  

In this case, coupled differential equations of pressure and density profile describe the envelope structure,
\beqar\label{polytrope}
{d \rho \over dr} &=  - {G M_1  \over r^2} {\rho^2 \over \Gamma_{\rm s} P} = - g  {\rho^2 \over \Gamma_{\rm s} P} \nonumber \\
{dP \over dr} &= - {G M_1  \over r^2} \rho  =  -g \rho
\eeqar
where $g =  {G M_1  / r^2} $. The parameter $\Gamma_{\rm s} = 1 + 1/n$ is the polytropic index of the stellar profile such that 
 \beq
\left( {d \ln P \over d \ln \rho} \right)_{\rm envelope}= \Gamma_{\rm s},
\eeq
where the subscript denotes that this expression is evaluated along the envelope profile -- a change in density in the stellar profile implies a change in pressure, $P_{1} \propto {\rho_{1}}^{\Gamma_{\rm s}}$.

\subsection{Gas Equation of State}

The envelope gas may have a different response to compression than its arrangement in the hydrostatic profile. 
General equations of state have four adiabatic indices, which describe their thermodynamic behavior \citep[see, for example, chapter 3 of][]{2004sipp.book.....H}, in these cases there may be departures between $\gamma_1$, defined by
\beq
\left( {d \ln P \over d \ln \rho} \right)_{\rm ad}= \gamma_1,
\eeq
and $\gamma_3$, which is defined by
\beq
\left( {d \ln T \over d \ln \rho} \right)_{\rm ad}= \gamma_3 - 1,
\eeq
where the subscript indicates partial derivatives along an adiabat (at constant entropy). The gas's adiabatic behavior is particularly relevant because any compression induced by a companion will happen on a timescale much shorter than the stellar envelope thermal timescale.  
The first exponent, $\gamma_1$, is relevant in computing the gas sound speed, $c_{\rm s}^2 = \gamma_1 P / \rho$, and the third, $\gamma_3$, enters into the equation of state relationship between pressure, density, and internal energy as $P=(\gamma_3 -1 ) \rho e$. 
Constant entropy stellar envelope structures (for example, a convective envelope) have $\gs \approx \gamma_1$, while other structures (like a radiative envelope) may have $\gs < \gamma_1$. 

For an ideal gas, all of the adiabatic exponents are identical. In this case,  there is a single adiabatic exponent, $\gamma$, which is
\beq\label{ideal}
\gamma = \gamma_1 = \gamma_3.
\eeq
This implies that when ideal gas is compressed (or allowed to expand) adiabatically, the pressure follows $P\propto \rho^\gamma$. 

\subsection{Relationships between Flow Parameters}\label{sec:rel}

In a common envelope encounter, the secondary star plunges into the envelope of the primary. We use the (simplified) polytropic description above to show that there are relationships between the characteristic flow parameters described in Section \ref{sec:scales}. 

These relationships directly result from the fact that the envelope is in hydrostatic equilibrium in opposition to the same gravitational forces that determine the secondary object's orbit. Therefore, these relationships will hold for any hydrostatic envelope structure, which needs not be the pre-encounter stellar envelope structure. 

To derive the relationships between the characteristic scales of common envelope flows, we start with the pressure gradient of the envelope, 
\beq
{d P \over dr}  = - g \rho.
\eeq
We re-write the left-hand side as $dP/dr = (dP/d\rho)\times(d\rho / dr)$.  
We can then use the derivative of pressure with respect to density within the envelope to find,
\beq
{d P \over d \rho} = \gs {P \over  \rho} = \gamma_1 \left({\gs \over \gamma_1}\right) {P \over  \rho} = \left({\gs \over \gamma_1}\right)  c_s^2,
\eeq
because $ c_s^2 = \gamma_1 {P /  \rho}$. 
Substituting this in, our expression becomes,
\beq
{c_s^2 \over \rho} {d\rho \over dr} = -g \left({\gs \over \gamma_1}\right)^{-1}.
\eeq
We substitute in the definition of the density scale height (equation \ref{hrho}), the definition of $g$, and set $r=a$ (the separation of the pair) to find
\beq
{c_s^2  \over H_\rho} = {G M_1 \over a^2} \left({\gs \over \gamma_1}\right)^{-1}.
\eeq
Further substitutions into this expression are useful. We use the definition of $\vk$, equation \eqref{vkdef}, and total system mass $M=M_1+M_2$ to write
\beq
c_s^2  H_\rho^{-1} = {\vk^2 \over a}{M_1 \over M} \left({\gs \over \gamma_1}\right)^{-1}.
\eeq
Then, rearranging and introducing the Mach number, equation \eqref{machdef}, based on a flow relative velocity $\vinf = \fk \vk$, 
\beq
  {a \over H_\rho} =  {\vk^2 \over c_s^2} {M_1 \over M} \left({\gs \over \gamma_1}\right)^{-1}  = {\mach^2 \over \fk^2} {M_1 \over M} \left({\gs \over \gamma_1}\right)^{-1}.
\eeq
We can then substitute in for the accretion radius $\Ra$ and $\erho$, equations \eqref{radef} and \eqref{erhodef}, to express the relationships between the flow parameters, 
\beq\label{mach}
\mach^2 =  \erho {(1+q)^2 \over 2q} \fk^4 \left({\gs \over \gamma_1}\right)
\eeq
or 
\beq\label{erho}
\erho = {2 q \over (1+q)^2} \mach^2 \fk^{-4} \left({\gs \over \gamma_1}\right)^{-1}.
\eeq
We note that where the enclosed primary-star mass, $m_1(a)$, is substantially less than $M_1$, the enclosed mass may be replaced into the above equations by using $q_{\rm enc} = M_2/m_1(a)$ in place of $q$. These expressions are extremely useful because they reduce the multi-parameter space of common envelope flows down to a plane of allowed combinations on the basis of the hydrostatic equilibrium nature of the stellar envelopes. 

\subsection{Example Profiles for Two Primary Stars}
\begin{figure*}[htbp]
\begin{center}
\includegraphics[width=0.8\textwidth]{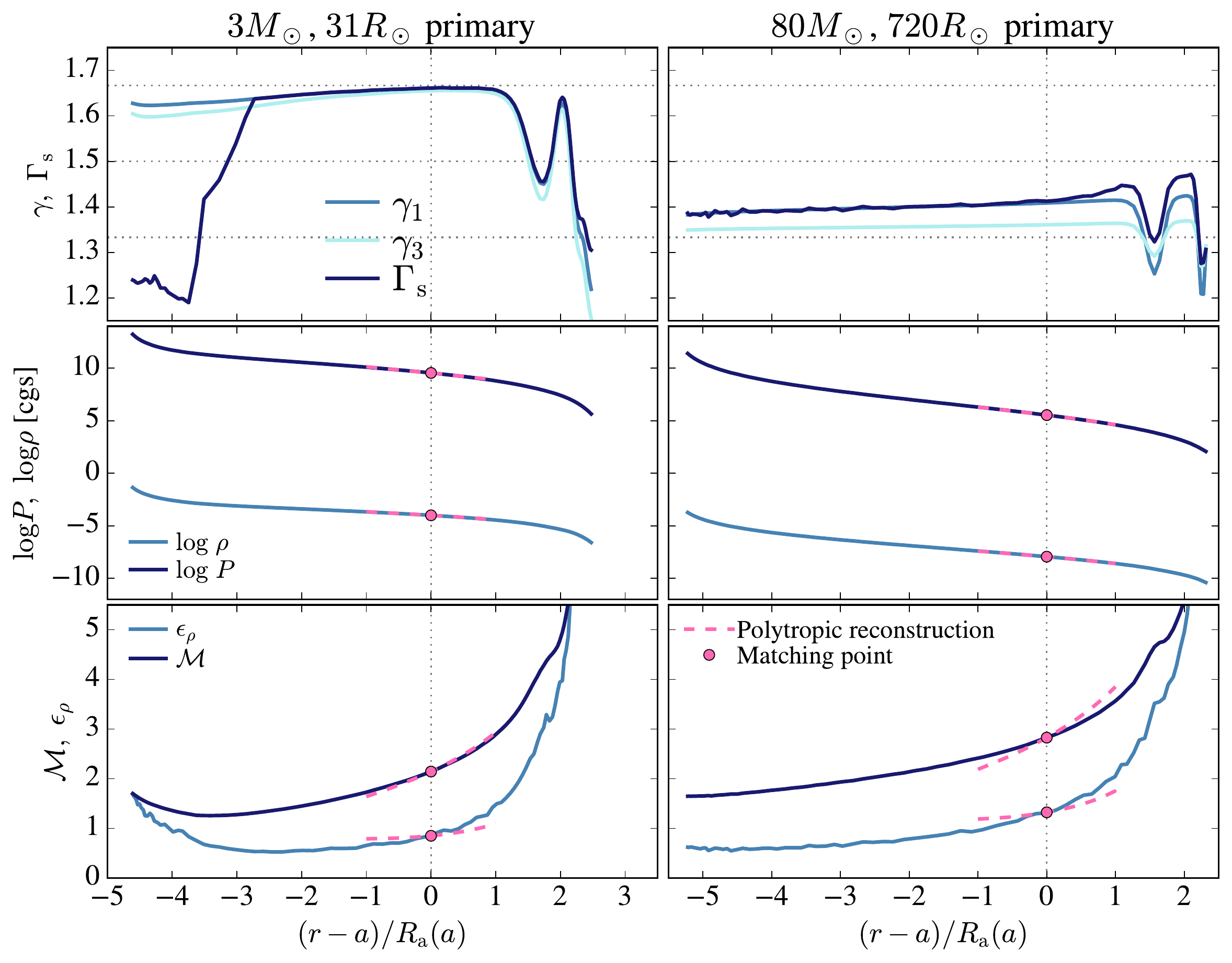}
\caption{Profiles of primary-star stellar structure relevant to common envelope inspiral. A secondary star is embedded within the envelope of the primary at the separation marked with the vertical line. The $x$-axis shows radial distance in units of the accretion radius, $\Ra$. The top panel compares profiles of gas adiabatic exponents, $\gamma_1$ and $\gamma_3$, along with the local structural parameter $\Gamma_{\rm s}$. Note that $\gamma_1$ and $\gamma_3$ are relatively similar to each other, and additionally, that in convective regions of the stellar envelopes $\gamma_1 \approx \gs$. The center panel shows profiles of density and pressure, with local polytropic reconstructions extending $\pm \Ra$ (pink dashed lines). The lower panel shows that these properties can be matched to a flow Mach number, $\mach$ and density gradient $\erho$, at the position of the embedded object. The slope of the polytropic profile of these secondary parameters is approximate but not perfectly fit, however, as can be seen in the lower panel.  }
\label{fig:stellarstructure}
\end{center}
\end{figure*}

We illustrate these relationships between flow parameters using a secondary object $(q=0.1)$ embedded within the unperturbed envelopes of two giant stars in Figure \ref{fig:stellarstructure}. 

To compute these envelope profiles, we use the MESA stellar evolution code, version 8845 \citep{2011ApJS..192....3P,2013ApJS..208....4P,2015ApJS..220...15P}. 
We show a $3M_\odot$ red giant with $Z=0.01$ that has evolved to have a $31R_\odot$ radius and an $0.43M_\odot$ helium core. The input list for this model is based on the \verb|7M_prems_to_AGB| test suite input list, with a change to $3M_\odot$. This input includes a mixing length $\alpha$ of 1.73, and a Reimer's red giant branch wind prescription with coefficient 0.5. We also show a lower metallicity massive star, which is $80M_\odot$ with $Z=0.001$ and has evolved to $720R_\odot$, with a $41.2M_\odot$ helium core. This model was run using the \verb|150M_z1m4_pre_ms_to_collapse| test suite example inlist, modified to $80M_\odot$ initial mass and $Z=0.001$, and no other modifications to the \verb|inlist_massive_defaults| parameters, which include a mixing length $\alpha$ of 1.5, and a semiconvection $\alpha$ of 0.01.\footnote{input lists available upon request to the corresponding author.} 

The panels of Figure \ref{fig:stellarstructure} map out profiles of gas compressibility, density,  and pressure within the stellar envelope  along with profiles of Mach number and density gradient. 
The panels are normalized to the location of a hypothetical secondary, embedded within the stellar envelope, and the x-axes show distance in units of the accretion radius of this object, $\Ra$. Within $\pm\Ra$, we also show a polytropic reconstruction of the local profile. These panels adopt $q=0.1$ (which implies secondary masses of $0.3M_\odot$ and $8M_\odot$, respectively) and $\fk=1$. 

The flow in these and other common envelope encounters is described by these profiles of pressure and density across the accretion radius -- but this description can be compactly represented in the parameters of the Mach number, density gradient, gas compressibility ($\gamma_1$, $\gamma_3$), and structural gamma ($\gs$). The panels of Figure \ref{fig:stellarstructure} show that equations \eqref{mach} and \eqref{erho} reproduce the Mach number and density gradient at a given position, and that a polytropic profile reproduces the approximate slope of these parameters around the central value. 

The two examples in Figure \ref{fig:stellarstructure} show overall similarity despite originating in relatively different stars: the highest Mach numbers and density gradients are found near the stellar limb where radiative losses contribute to a reduction of the scale height. Typical Mach numbers are $\mach \sim 1-5$ and density gradients, $\erho$,  are of order unity. Gas adiabatic exponents, $\gamma_1$ and $\gamma_3$, and the structural parameter, $\Gamma_{\rm s}$, are both $\approx 5/3$ in the interior of the $3M_\odot$ star's convective envelope, but drop to lower values in zones of partial ionization nearer to the surface. The values of $\gamma_1$ and $\gamma_3$  diverge from $\gs$ in the radiative interior of the $3M_\odot$ star. The more massive, $80M_\odot$, star has an extended convective envelope (so $\gs\approx\gamma_1$) and a more compressible equation of state, with $\gamma_1 \sim1.4$ and $\gamma_3 \sim 1.35$ due to a partial contribution to the pressure from radiation \citep[see, e.g.][for more details]{2017A&A...597A..71S}. 

\section{Numerical Approach: Common Envelope Wind Tunnel}\label{sec:numericalmethods}

We study flows under typical common envelope conditions using an idealized three-dimensional hydrodynamic setup, which we call the Common Envelope Wind Tunnel. This section outlines the details of our numerical method.

\subsection{Hydrodynamic Implementation}
We solve the equations of inviscid hydrodynamics using the FLASH code \citep{2000ApJS..131..273F}. FLASH is a grid based code with adaptive mesh refinement. We use the directionally split Piecewise Parabolic Method Riemann solver in the calculations presented here \citep{1984JCoPh..54..174C}. We use an ideal gas, gamma-law equation of state, 
\beq
P=\left(\g -1 \right) \rho e
\eeq
and take different values of the compressibility, $\g$ in different simulations. As noted in equation \eqref{ideal}, the ideal gas approximation assumes that $\g=\gamma_1=\gamma_3$, which is generally a reasonable (but inexact) approximation for  thermodynamic conditions of interest for stellar envelopes -- see Figure \ref{fig:stellarstructure} to note the small departures between $\gamma_1$ and $\gamma_3$ in the MESA stellar models due to their more sophisticated treatment of the equation of state. 

Like the simulations of \citet{2015ApJ...803...41M}, our Common Envelope Wind Tunnel calculations have a 3D cartesian computational domain with a point mass representing the embedded object at the coordinate origin. The simulations are performed in dimensionless units, in which $\Ra = \vinf = \rhoinf = 1$, where $\rhoinf$ is the density of the primary-star envelope at $r=a$. These units imply a time unit of $R_a/\vinf=1$, or one flow crossing time of the accretion radius. The mass of the embedded object is $M_2=(2G)^{-1}$ in these simulation units, and the primary has mass $M_1 = q^{-1} M_2$. Because $\vinf = \fk \vk = 1$, we can solve for the binary separation, $a$ in code units,
\beq
a = \fk^2 G M = {1\over 2} \fk^2 (1+q^{-1}).
\eeq
The orbital plane defined in the simulation is the $x-y$ plane. We locate the primary at $y_1 = -a$. The gravitational force from the primary therefore acts in the $-y$ direction. As a concession to the cartesian geometry of our domain, the primary-star gravity only depends on the $y$-coordinate,
\beq\label{agrav1}
\vec{a}_{\rm grav,1} = - \frac{G M_1}{(y-y_1)^2} \hat{y}
\eeq
where $\vec{a}_{\rm grav,1}$ is the gravitational acceleration from the primary star's gravity. The acceleration from the point mass at the coordinate origin (representing $M_2$) is
\beq\label{agrav2}
\vec{a}_{\rm grav,2} = - \frac{G M_2}{|\vec{r}|^2} \hat{r}.
\eeq
where $\vec{r}=(x,y,z)$ is the distance from the coordinate origin  to the cell, where we are calculating the force within the computational domain.

The simulation $-x$ boundary feeds a wind into the wind tunnel and past the point mass. The wind has a gradient of pressure and density in the $y$ direction, and is uniform in the $z$ direction.  The conditions of this wind are specified by an upstream Mach number, $\mach$, a density gradient, $\erho$, and the pressure and density at $y=0$ (along the $x$ axis).  We begin by specifying a corresponding pair of $\erho$ and $\mach$ given $q$, $\fk$, $\gs$, and $\gamma$ using equations \eqref{mach} and \eqref{erho}.  We then assign $P_\infty$ to generate the sound speed that satisfies the selected flow Mach number, based on $\rhoinf$ and $c_s^2 = \gamma P / \rho$. We therefore have $P_\infty = \mach^{-2}   \gamma^{-1} \rhoinf \vinf^2$ (note that $\rhoinf = \vinf=1$ in our code units).  

Once the values at $y=0$ are set, we integrate to both positive and negative $y$ using the expressions of hydrostatic equilibrium for a mass-less atmosphere, see equation \eqref{polytrope}. Here the relevant differentials in our code units become $dP/dy$ and $d\rho/dy$.  We extend the hydrostatic profile to the ghost zones in the $-y$ boundary such that the hydrostatic pressure gradient is preserved.  The wind fed into the box is therefore in hydrostatic equilibrium with the primary-star gravitational force, and since it is supported on its lower boundary, it does not rise or fall unless an additional force is applied. On the $+x$, $+y$, and $\pm z$ boundaries, we apply `diode' boundary conditions, which allow material to freely leave but not enter the grid. 

The initial condition is uniform in the $x$ and $z$ directions with flow properties based on the integrated profile of pressure and density in the $y$ direction. The velocity everywhere is set to $\vec{v} = \vinf \hat{x}$. We turn the central point mass on progressively over the first code time unit, so $\vec{a}_{\rm grav,2}$ is fully active for $t>1 \Ra/\vinf$. 

We create an absorbing central ``sink'' surrounding the point mass, with radius $\Rs$.  The calculations in this paper use $\Rs=0.05\Ra$. Each timestep, the average pressure and density of a spherical shell, which extends from $\Rs$ to $2\Rs$ are computed. The conditions inside the sink are reset to a fraction, usually $10^{-3}$ of these values, creating an effective vacuum -- and deleting (accreting) mass and energy from the grid every timestep. This prescription represents accretion without feedback on the surrounding flow \citep{2015ApJ...803...41M}.

Our computational domain extends from $\pm 4 \Ra$ in the $x$ direction and $\pm 3.5 \Ra$ in the $y$ and $z$ directions. This domain is covered by 8 blocks in $x$ and 7 each in $y$ and $z$ of $8^3$ cells in each direction. We employ the PARAMESH adaptive mesh refinement package (v4), and base refinement choices on the second derivative of gas internal energy (erg g$^{-1}$) \citep{2000CoPhC.126..330M}.  We set the minimum refinement level to 2 (so all blocks are refined at least once) and the maximum refinement to 6. The maximum cell size is therefore $\Ra/16$ and the minimum is $\Ra/256$. As in \citet{2015ApJ...798L..19M}, to focus the highest resolution cells in the center of the computational domain (near $M_2$), we drop the maximum refinement with distance. Blocks with size less than $\alpha r$ (we adopt $\alpha=0.3$) are not allowed to refine further. The first drop in refinement occurs at $\gtrsim1.5 \Rs$ and drops one level each time the distance from the point mass doubles.

\subsection{Diagnostics of Flow Properties}\label{sec:diag}
Several key diagnostics and integral quantities of the flow are computed at runtime and recorded every timestep in our setup. 

\subsubsection{Accreted Quantities}
We record the properties of material that falls into the central sink just prior to deleting it. These quantities represent the accreted mass, angular momentum, and linear momentum. Each timestep, we perform a volume integral over the sink cells and sum the total accreted quantity. We convert this sum to an accretion rate by dividing by the timestep. For example,
the accretion rate of mass is given by 
\beq\label{mdotdiag}
\dot M = {1\over \Delta t} \int_{\rm sink} (\rho - \rho_{\rm sink}) dV,
\eeq
where $\rho_{\rm sink}$ is the density the sink cells were set to on the previous timestep. Similarly,  
\beq\label{pdotdiag}
\dot p_x = {1\over \Delta t} \int_{\rm sink} (\rho - \rho_{\rm sink}) v_{x} dV.
\eeq
is the accretion rate of linear momentum along the direction of motion. This accretion of momentum also represents a force that we will call $F_{\dot p_x}$ in what follows. 

\subsubsection{Gravitational (Dynamical Friction) Drag Forces}
If the mass distribution around the embedded object is not spherically symmetric, it experiences a net gravitational force. The component of this force directed along the direction of motion of the object constitutes a ``gravitational drag'' or gas dynamical friction that modifies the motion of the gravitating object \citep{1943ApJ....97..255C,1999ApJ...513..252O}. In the case of supersonic flows past a gravitating object, an overdense wake is generated that exerts a stronger gravitational force than the upstream gas \citep{1999ApJ...513..252O}. The net force decelerates motion, and is thus termed a drag. 

The gravitational force on the secondary, $M_2$ by a volume of gas $dV$ is 
\beq
d\vec{F}_{\rm grav} = \frac{G M_2 \rho dV}{r^2} \hat{r},
\eeq
(note the inversion of the sign in this expression as compared to equation \ref{agrav2}). The component of this force along the direction of motion is,
\beq
dF_{{\rm grav},x} = \frac{G M_2 \rho  dV x}{r^3}.
\eeq
The net dynamical friction drag, $F_{\rm df}$, is the volume integral of the contributions to the gravitational force along the direction of motion $dF_{{\rm grav},x}$, 
\beq\label{fdf}
F_{\rm df} = \int dF_{{\rm grav},x}.
\eeq
The sign of $F_{\rm df}$ in our coordinate setup is such that a positive value represents a drag force (deceleration).\footnote{Alternatively, one could also calculate the drag force by measuring the momentum  and pressure change of the gas (rather than the gravitational net force on the particle).  We note here that  \citet[][equation 3]{2008ApJ...672L..41R} and \citet[][equation 13]{2015ApJ...803...41M} used this approach and measured the net momentum transport by gas passing through a spherical surface to evaluate drag forces generated within the enclosed volume. However, in their analysis of the momentum equation in steady state, \citet{2016A&A...589A..10T} show that these expressions are incomplete because they do not include the difference in  pressure across the surface. In their section 3.5, \citet{2016A&A...589A..10T} show that the sum of a surface integral of momentum transport and a surface integral of (net) pressure balance dynamical friction (see \citet{2016A&A...589A..10T}, section 3.5 for a complete derivation).  For this application, that of supersonic Hoyle-Lyttleton flow, the pressure term is opposite in sign and smaller in magnitude than the momentum transport term \citep[see, for example, Figure 3 of][]{2016A&A...589A..10T}.  This suggests that the drag forces derived from \citet[][equation 3]{2008ApJ...672L..41R} and \citet[][equation 13]{2015ApJ...803...41M} are of the correct magnitude, but are likely moderate overestimates of the drag force generated within a particular volume. 
}

\section{Numerical Results}\label{sec:numericalresults}

In this section, we present and analyze two suites of calculations, each spanning a range of density gradients and corresponding Mach numbers. Each calculation adopts $\Rs=0.05\Ra$. In one suite, we take $\g=\gs=5/3$, relevant for the convective envelopes of low-mass stars, as shown in Figure \ref{fig:stellarstructure}. In a second suite, we take $\g=\gs=4/3$, as exemplifying the high-compressibility limit of massive star envelopes in which radiation pressure becomes increasingly important. Here we compare some key flow properties, including rates of accretion and the generation of drag forces, realized in these simulations.

\subsection{Flow Morphology}\label{sec:flow}

\begin{figure*}[htbp]
\begin{center}
\includegraphics[width=0.95\textwidth]{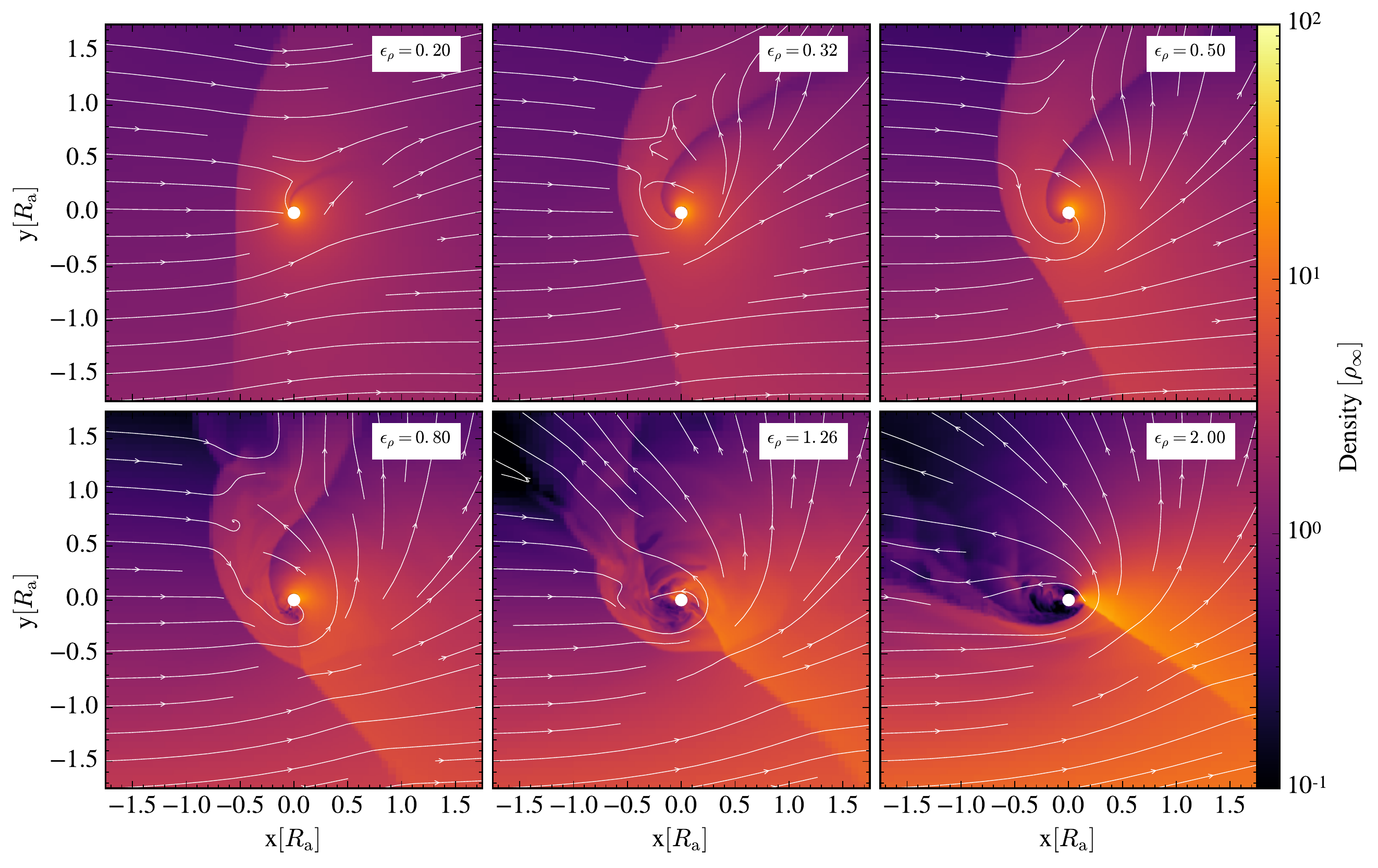}
\includegraphics[width=0.95\textwidth]{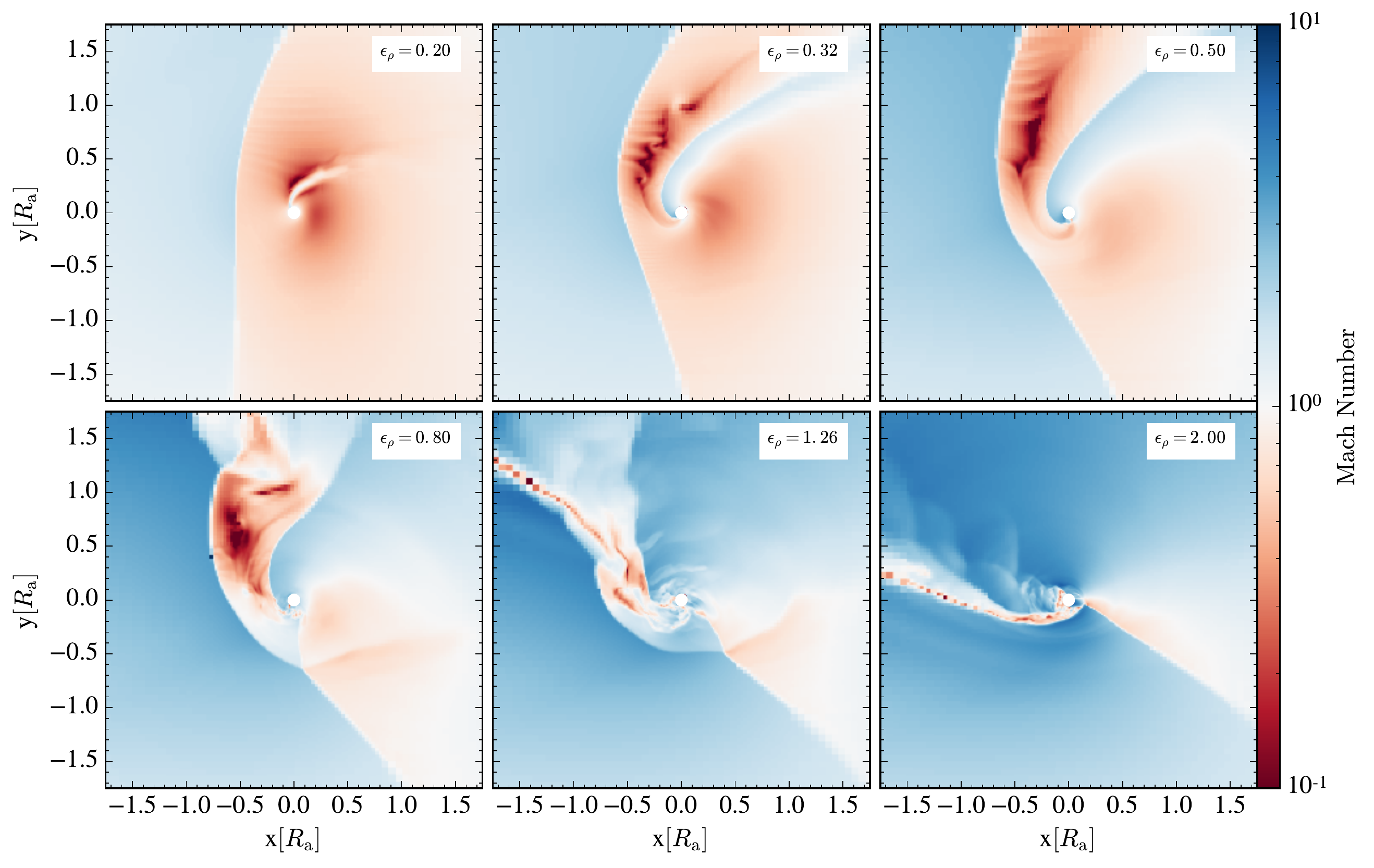}
\caption{Slices through the $z=0$ (orbital) plane surrounding an object embedded in the common envelope wind tunnel with $\gamma=\Gamma_{\rm s}=5/3$. The snapshots compare flow at $t=20\Ra/\vinf$. The upper panels show density in units of $\rhoinf$, while the lower panels show Mach number.  }
\label{fig:g53z}
\end{center}
\end{figure*}

\begin{figure*}[htbp]
\begin{center}
\includegraphics[width=0.95\textwidth]{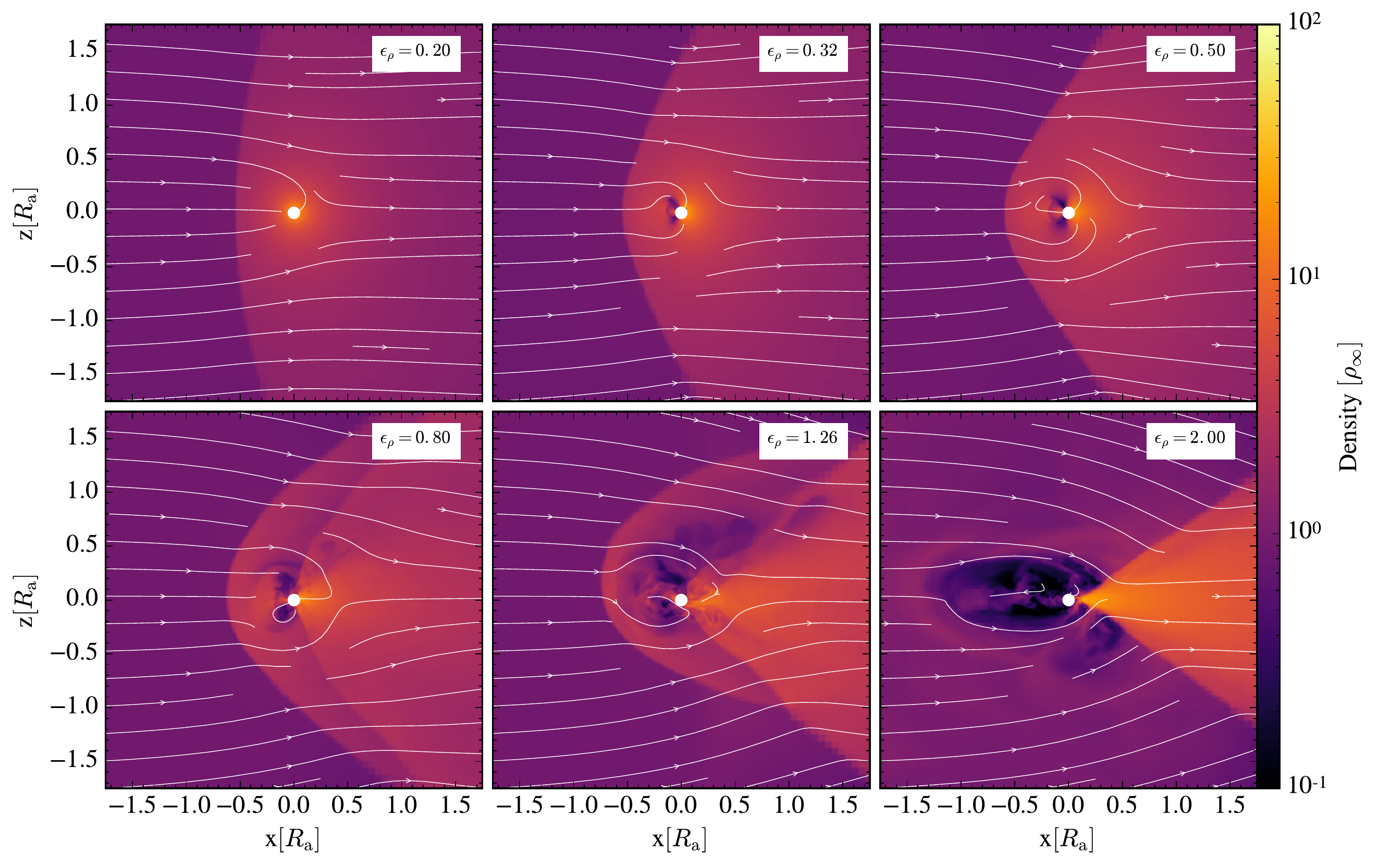}
\includegraphics[width=0.95\textwidth]{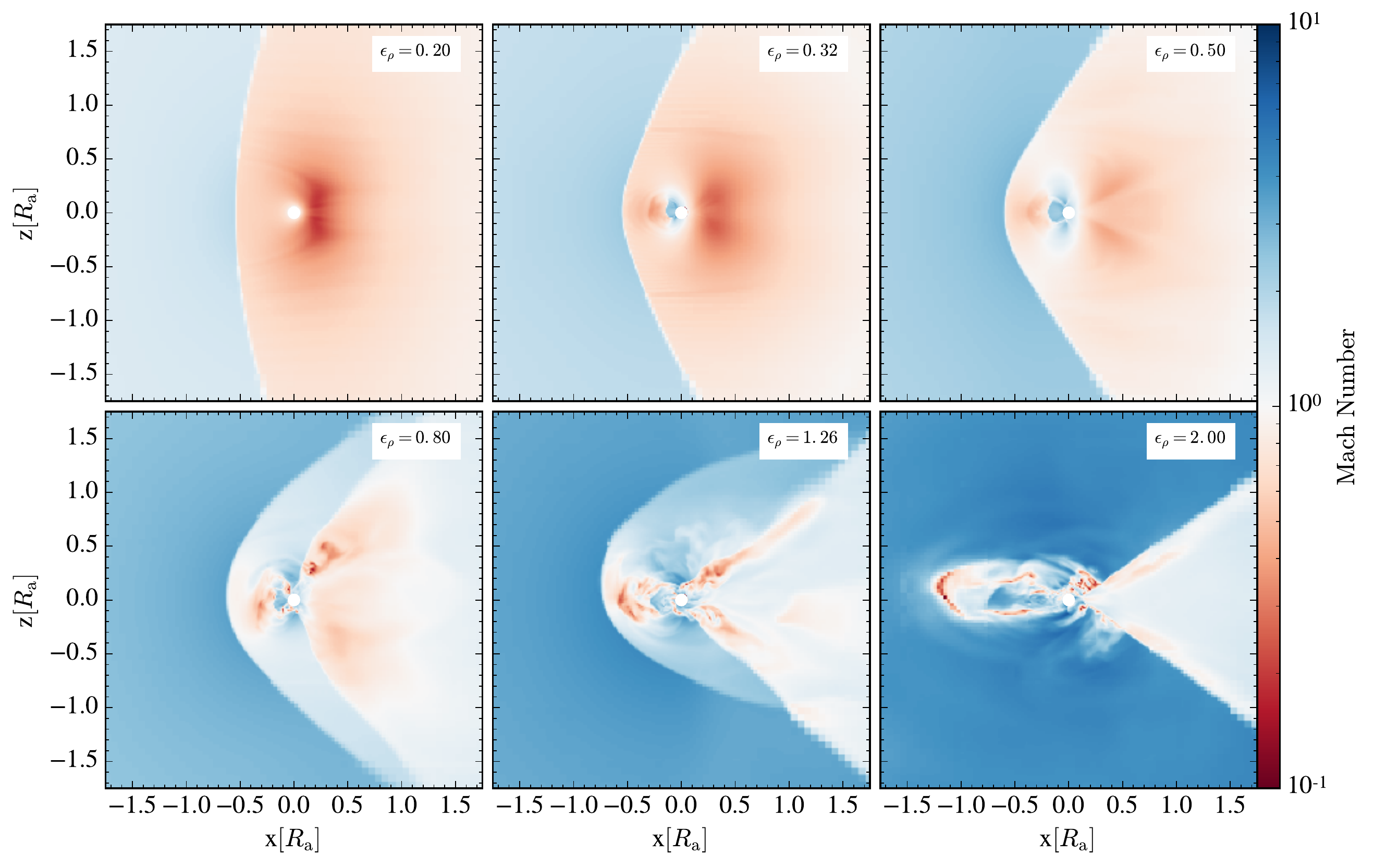}
\caption{Slices through the $y=0$ (perpendicular to orbital) plane surrounding an object embedded in the common envelope wind tunnel with $\gamma=\Gamma_{\rm s}=5/3$. The snapshots compare flow at $t=20\Ra/\vinf$. The upper panels show density in units of $\rhoinf$, while the lower panels show Mach number.  }
\label{fig:g53y}
\end{center}
\end{figure*}

\begin{figure*}[htbp]
\begin{center}
\includegraphics[width=0.95\textwidth]{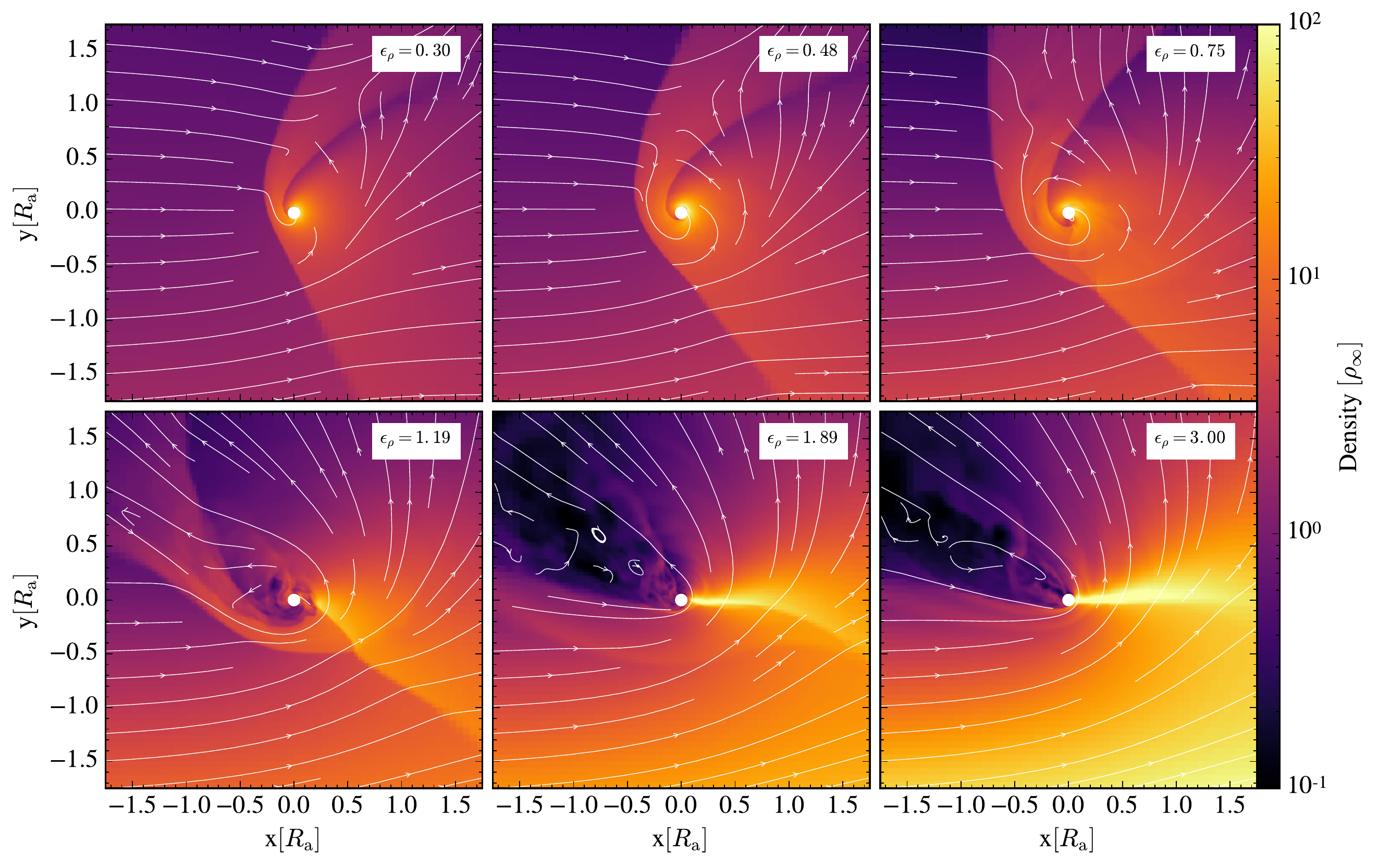}
\includegraphics[width=0.95\textwidth]{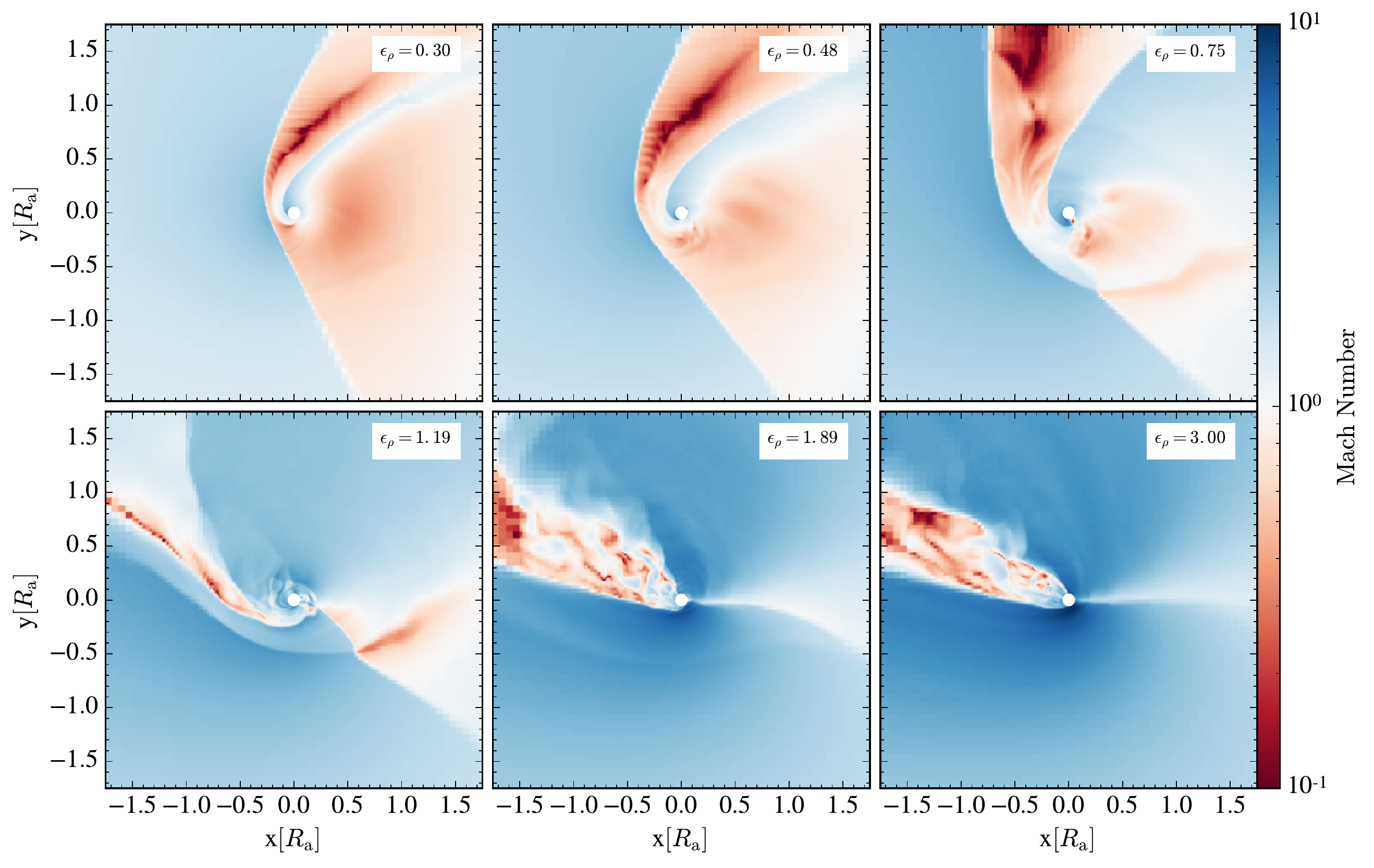}
\caption{Same as Figure \ref{fig:g53z} for $\gamma=\Gamma_{\rm s}=4/3$. Slice in the orbital plane. }
\label{fig:g43z}
\end{center}
\end{figure*}

\begin{figure*}[htbp]
\begin{center}
\includegraphics[width=0.95\textwidth]{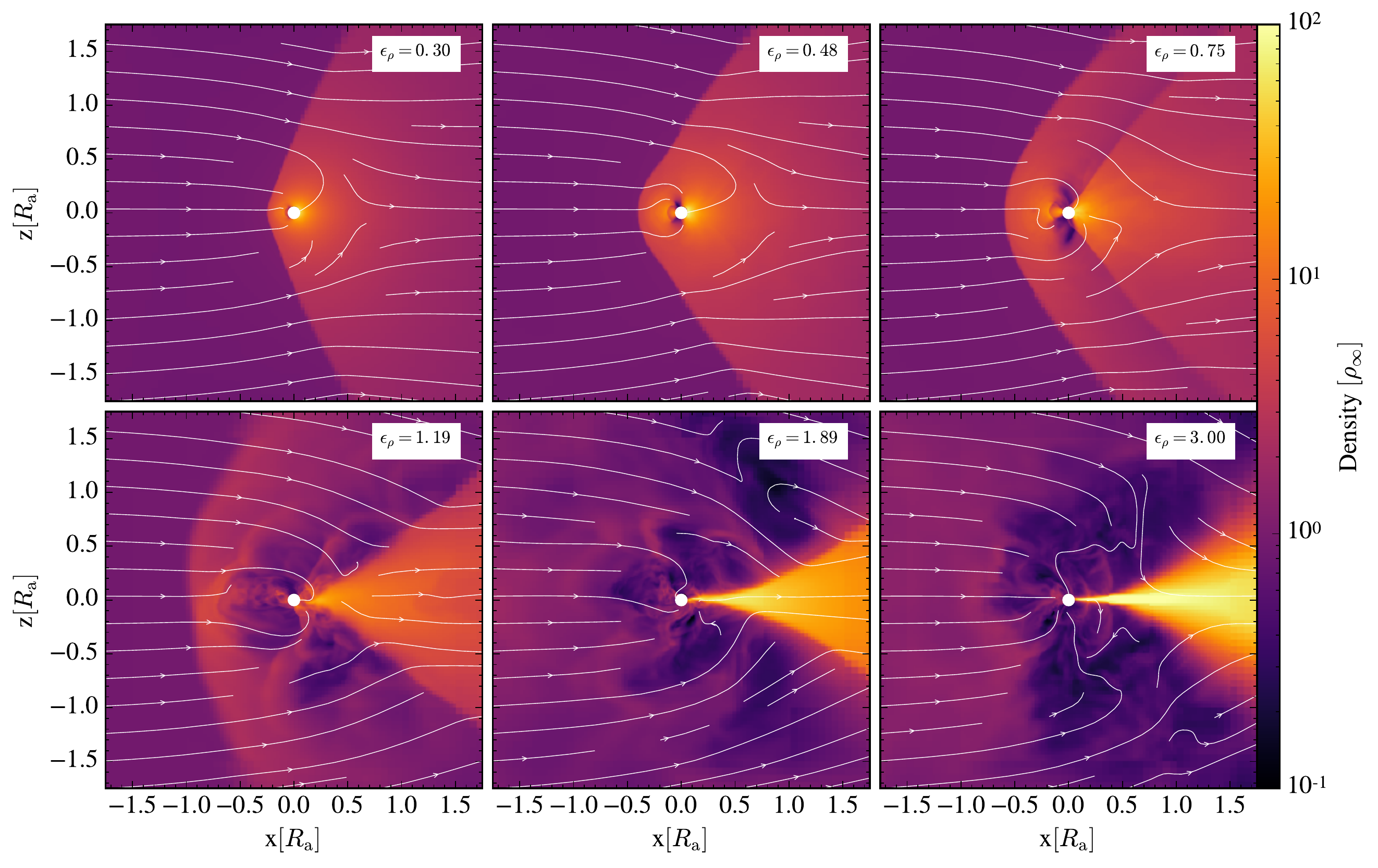}
\includegraphics[width=0.95\textwidth]{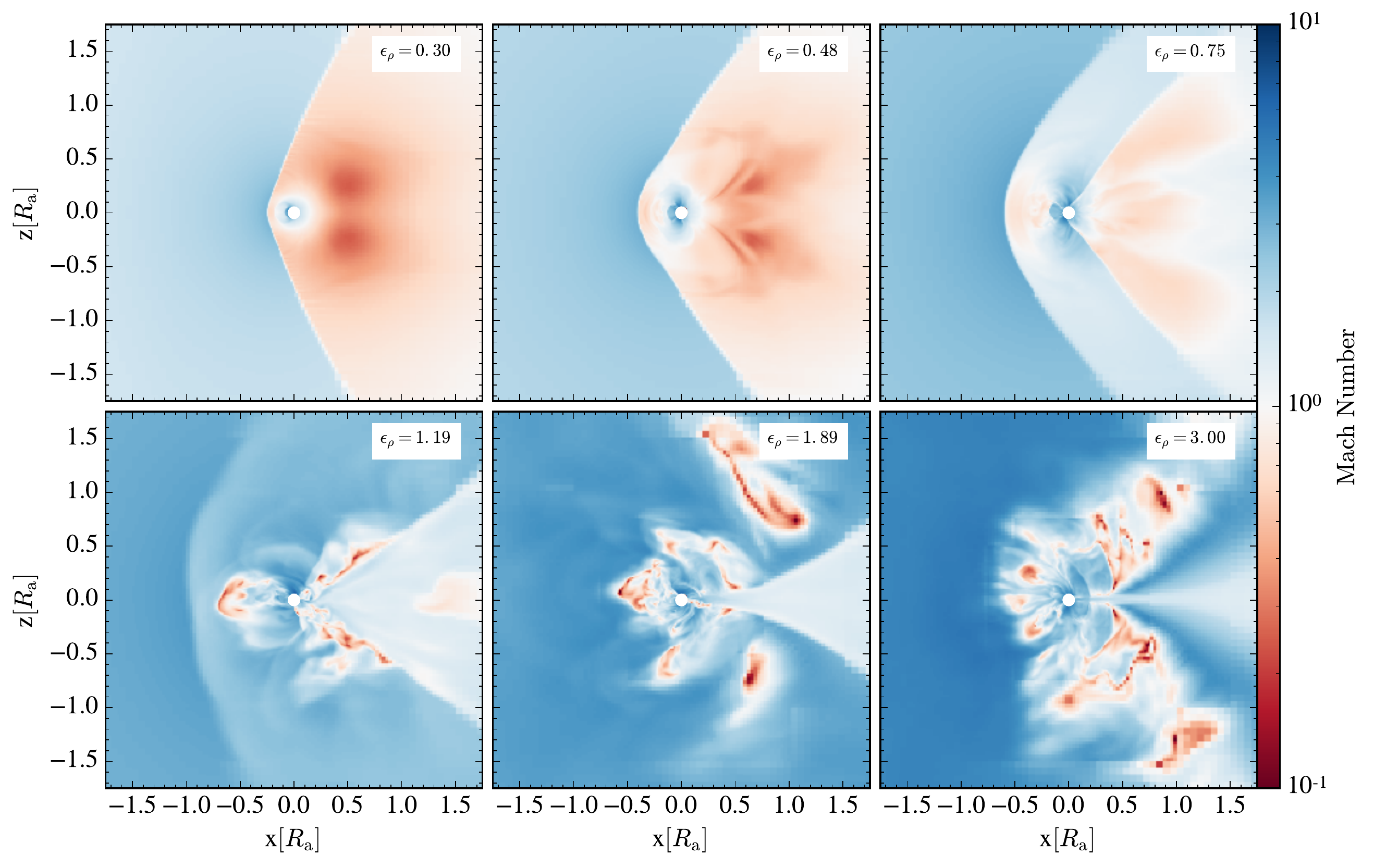}
\caption{Same as Figure \ref{fig:g53y} for $\gamma=\Gamma_{\rm s}=4/3$. Slice perpendicular to the orbital plane.}
\label{fig:g43y}
\end{center}
\end{figure*}

The morphology of flows around objects embedded in the common envelope is distorted and asymmetric in response to the gradient of density in the upstream, stellar envelope material \citep{2015ApJ...803...41M}. Our new calculations are consistent with this result, but the inclusion of the primary-star properties -- in the form of gravitational force, hydrostatic equilibrium pressure gradients, and the relationships between flow parameters discussed in Section \ref{sec:rel} -- impacts the expected nature of common envelope flows.  

In Figure \ref{fig:g53z} through Figure \ref{fig:g43y}, we show slices of density and Mach number through the orbital ($x$-$y$) plane, and perpendicular to the orbital $(x-z)$ plane for the simulation suites with $\gamma=\Gamma_{\rm s}=5/3$ and $\gamma=\Gamma_{\rm s}=4/3$, respectively.  In the each case, we run simulations for six log-spaced values of the density gradient,  $\erho$. For $\gamma=\Gamma_{\rm s}=5/3$, this corresponds to $\erho= 0.2, 0.32, 0.5, 0.8, 1.26, 2.0$. For $\gamma=\Gamma_{\rm s}=4/3$, we have $\erho= 0.3, 0.48, 0.75, 1.19, 1.89, 3.0$. The ranges of density gradient parameters were chosen such that in the steepest gradient cases, the object was embedded to a depth of approximately one accretion radius within the envelope of the primary.  

The corresponding upstream Mach numbers also vary across these simulations, following equation \eqref{mach} with $\fk=1$, from $\mach=1.1$ for $\erho=0.2$ to $\mach \approx 3.48$ for $\erho=2$ and $\mach\approx4.26$ for $\erho=3$. By comparison to Figure \ref{fig:stellarstructure} and the associated discussion, we can see that the steepest density gradients and the highest Mach numbers correspond to flow near the limb of the primary-star envelope, while shallower gradients (and lower Mach numbers) are found deep within the common envelope. Therefore, in our panels of Figures \ref{fig:g53z} through \ref{fig:g43y}, the upper left panels correspond to flow around a deeply embedded object, while the lower right panels correspond to flow around an object nearer to the envelope limb. 

The flow Mach number is plotted in the lower panel sets of Figures \ref{fig:g53z} through \ref{fig:g43y}. A dramatic transition occurs here with upstream density gradient. In the case of shallow density gradients, the symmetry of the bow shock is nearly preserved. Upstream flow is supersonic, while downstream flow is subsonic after crossing the shock and meeting a pressure gradient imposed by the convergence of flow into the post-shock region. As the density gradient steepens the portion of material in the post-shock region that remains supersonic increases dramatically. In flows with the steepest density gradients, the material moves nearly ballistically with only a small fraction having $\mach \ll 1$. 

One implication of the changing Mach number can be seen in the bow shock morphology. Bow shocks in homogenous flow exhibit an opening angle, $\Theta$, proportional to the Mach number, where $\sin \Theta \sim \mach^{-1} = \cs / \vinf$, because the disturbance from the shock wave moves laterally at approximately the sound speed while the stream motion is supersonic with $\vinf$. As the Mach number (and density gradient) increases in these simulations we see a narrowing of shock opening angles -- particularly along the well-defined edge facing the stellar center (and the flux of the densest material).  While the shock for $\erho=0.2$ in Figures \ref{fig:g53z} and \ref{fig:g53y}   is nearly planar, by the time $\erho=2.0$, the trailing shock opens in a much narrower cone. This consequence of choosing corresponding combinations of density gradient and Mach number can be contrasted to the simulations of \citet{2015ApJ...803...41M}, which adopted $\mach =2$ for all density gradients. 

The shock morphology shows an interesting secondary effect perpendicular to the orbital plane in cases of the steepest density gradient, particularly for $\erho = 1.19, 1.89, 3.0$ in Figure \ref{fig:g43y}.  In these snapshots, we see that the shock opening angle is not constant, but, in fact, widens with increasing displacement into the wake. What we observe from the streamlines in Figure \ref{fig:g43z} is that material focused onto the wake at larger $+x$ displacements comes from a larger impact parameter in the $-y$-direction. Recalling the profiles of Figure \ref{fig:stellarstructure}, this material, originating from deeper within the stellar envelope, has higher sound speed. As a result, there is a gradient of upstream Mach number in the $y$-direction (which can be observed in the lower panels of Figure \ref{fig:g53z} and \ref{fig:g43z}). The shock opening angle, which depends inversely on this upstream Mach number, thus broadens as the focussed material is drawn from deeper in the stellar envelope potential well. This effect is observable primarily in cases of steep gradient (near the envelope limb), where the derivatives of $\mach$ and $\erho$ become large.

The equation of state of the stellar envelope gas also plays a role in determining flow structure. 
The flow in the $\gamma=\Gamma_{\rm s}=4/3$ shown in Figures \ref{fig:g43z} and \ref{fig:g43y} is more compressible than the flow in the $\gamma=\Gamma_{\rm s}=5/3$ shown in Figures \ref{fig:g53z} and  \ref{fig:g53y}. This results in higher densities in the immediate wake of the embedded object because the pressure does not build up as rapidly upon compression in the focused material. In the steeper-gradient cases of $\gamma=\Gamma_{\rm s}=4/3$, we see a nested shock outside of an accretion line, which differs from the much broader fan of material in the $\erho=2$, $\gamma=\Gamma_{\rm s}=5/3$ simulation. 

In all cases, the secondary's gravitational focus lifts some dense material from the stellar interior against the primary star's gravity. 
This gravitational force leads some material  (with impact parameter $\gg \Ra$) to rise and fall in a ``tidal bulge'' trailing the embedded object. In material with impact parameter $\lesssim \Ra$, as shown in the streamlines overplotted on the upper panels of  Figures  \ref{fig:g53z} and \ref{fig:g43z}, this gravitational force leads to a slingshot around the embedded object. Some of this gas leaves the simulation box after being deviated through a large angle then expelled toward the lower-density of the primary star's limb ($+y$-direction in our simulation setup).

\subsection{Rates of Accretion}

\begin{figure}[tbp]
\begin{center}
\includegraphics[width=0.48\textwidth]{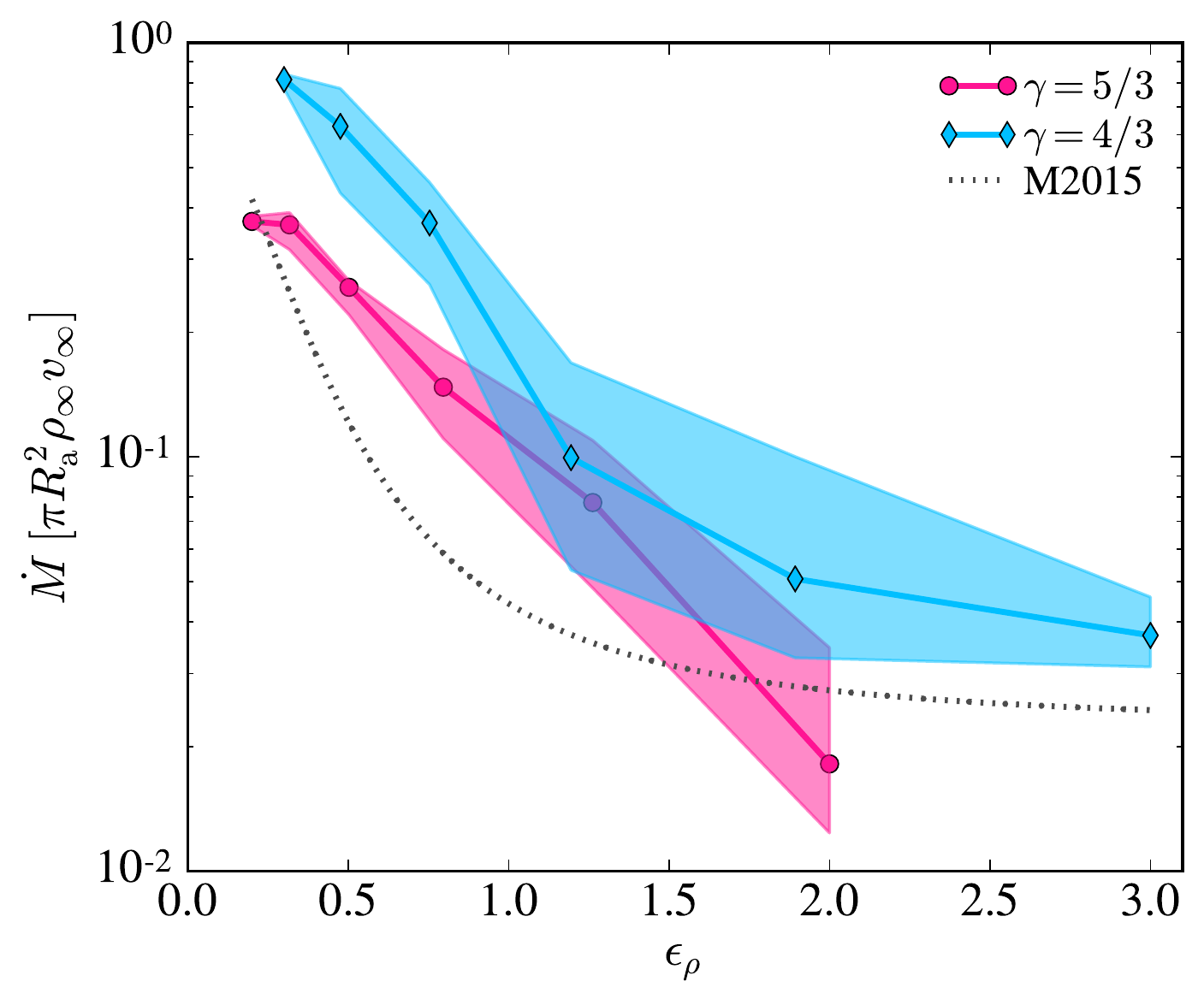}
\caption{Median mass accretion rates into the sink boundary condition defined by $\Rs=0.05\Ra$. Shaded regions denote the 5-th to 95-th percentile values of the time-variable $\dot M$. These are compared to the $\gamma=5/3$ case result of \citet{2015ApJ...803...41M}, which adopted $\mach=2$ for all simulations (labeled M2015).  In all cases, we find that steepening density gradient inhibits accretion, with typical values for large $\erho$ of $\dot M\ll \mhl$. The $\gamma=4/3$ cases show systematically higher  $\dot M$ than $\gamma=5/3$, perhaps because pressure gradients provide less resistance to flow convergence and accretion in the more compressible flow.  }
\label{fig:mdot}
\end{center}
\end{figure}

Our numerical approach replaces the embedded object with a sink on the grid, which absorbs convergent flow. The sink has a radius of $\Rs=0.05\Ra$. We note that this sink could be of a similar scale to that of a main-sequence star embedded in a typical common envelope \citep[see Table 1 of][]{2015ApJ...803...41M}, but is certainly much larger than the size of an embedded compact object like a white dwarf, neutron star, or black hole. Here we study rates and properties of material accreting through this inner boundary of our computational domain, but note that the accretion rate is dependent on the size of the sink boundary compared to the accretion radius \citep[for example,][found lower accretion rates for $\Rs=0.01\Ra$ than for $\Rs=0.05\Ra$]{2015ApJ...803...41M}. 

We begin by examining the mass accretion rate into the sink boundary as a function of density gradient in our $\gamma=\Gamma_{\rm s}=4/3$ and $\gamma=\Gamma_{\rm s}=5/3$ simulation suites, shown in Figure \ref{fig:mdot}.
Accretion rates in Figure \ref{fig:mdot} are normalized to the Hoyle-Lyttleton accretion rate,
\beq
\mhl = \pi \Ra^2 \rhoinf \vinf,
\eeq
which is the flux of material passing through a cross-section of area $\pi \Ra^2$ assuming a uniform density background. One feature of the accretion rate is that when density gradients are introduced into the flow, the flow morphology becomes less laminar and variability is introduced into the mass accretion rate. Therefore, we plot the median values (pink and blue lines) along with the 5-th and 95-th percentile ranges (shaded regions) for mass accretion rate, $\dot M$, as a function of density gradient, $\erho$. 

As density gradients steepen, the accretion rate into the sink drops dramatically and becomes more variable.    We see accretion coefficients ($\dot M/\mhl$) spanning more than an order of magnitude as density gradient changes across typical values. In all regions, the accretion efficiency is substantially lower than accretion from a uniform medium.  The imposition of a density gradient breaks the symmetry of the inflowing material (as seen in Figures \ref{fig:g53z} and \ref{fig:g43z}). As opposed to the uniform medium case, where momenta of opposing streamlines cancel, there is net angular momentum in the flow, which forms a barrier to efficient accretion when the circularization radius is significantly outside the sink radius \citep{2015ApJ...803...41M}. The increased  variability in cases of steep density gradient  can be attributed to the increased turbulence of the post-shock regions, as seen in Figures \ref{fig:g53z} through \ref{fig:g43y}. The more compressible $\gamma=4/3$ flow accretes at higher rates, particularly in cases of mild density gradient, $\erho\lesssim 1$, where radial pressure gradients oppose flow convergence less strongly than in the $\gamma=5/3$ case. 

Figure \ref{fig:mdot} also compares our new accretion rates to a fitting formula to the results of \citet{2015ApJ...803...41M} for $\Rs=0.05$, labeled M2015. The M2015 simulations all used $\mach=2$, $\gamma=5/3$, and the density gradient in the background material, $\epsilon_\rho$, was uniform rather than polytropic. Finally, there was no corresponding pressure gradient (a uniform pressure background was assumed). We find that despite these differences, accretion rates of similar order of magnitude are found. However, differences appear to lie in the functional form of $\dot M(\erho)$ and in the accretion rate for mild values of $\erho \lesssim 1$.  The two simulation suites presented here show higher mass accretion rates for $0.2 \lesssim \erho \lesssim 1.5$ by a factor of a few than M2015. One likely contribution to this difference is the lower Mach number in our current simulations for these density gradients. 

\begin{figure}[tbp]
\begin{center}
\includegraphics[width=0.48\textwidth]{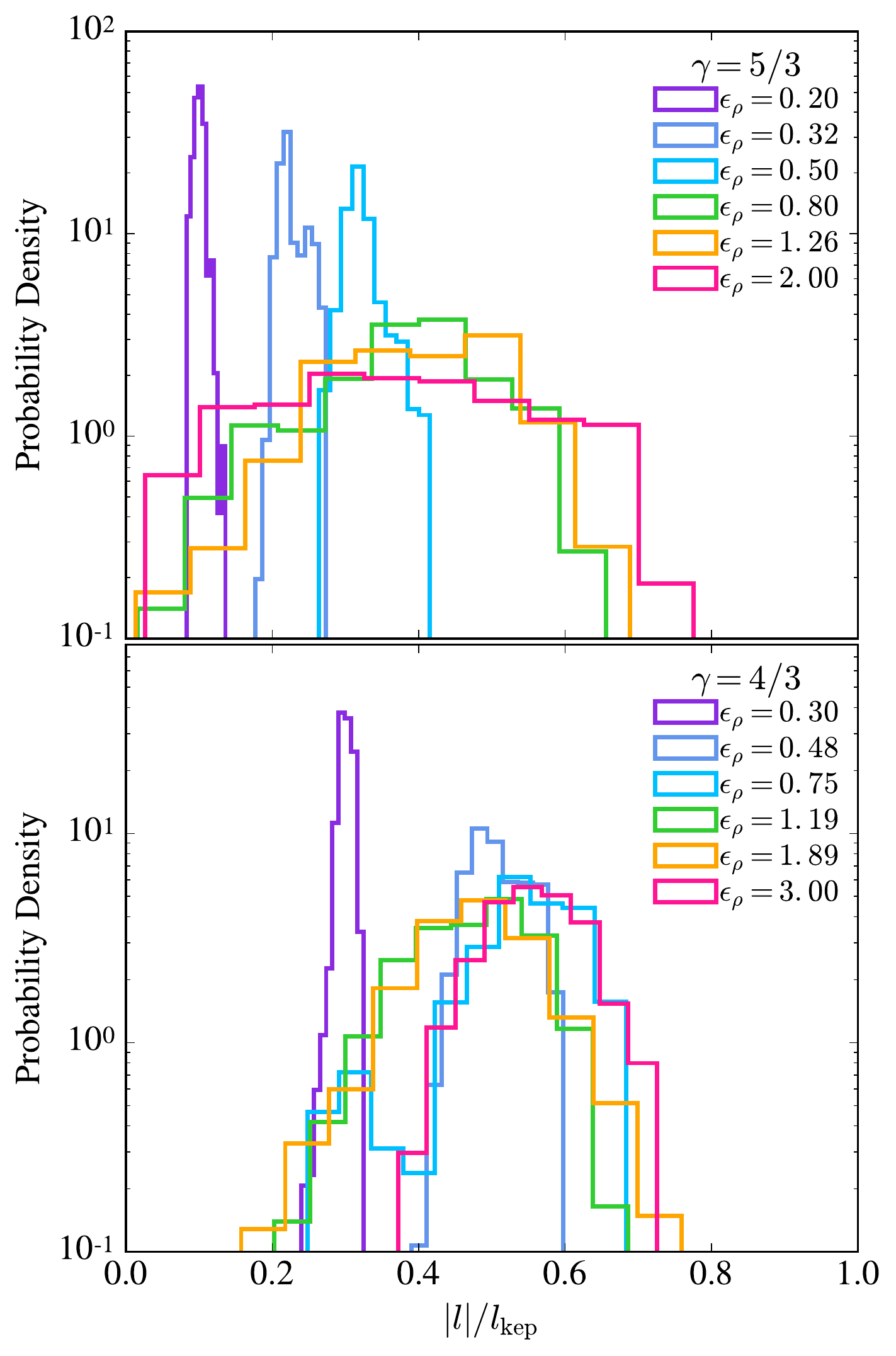}
\caption{Distributions of specific angular momentum of material accreted by the sink boundary condition ($\Rs=0.05\Ra$). Values are normalized to the specific keplerian angular momentum at the sink boundary: $l_{\rm kep} = \Rs v_{\rm kep}$. The distributions contain a range of $|l|/ l_{\rm kep}<1$, because material with full rotational support at the sink boundary $|l|/ l_{\rm kep}\gtrsim1$ would not accrete. In cases of a shallow density gradient, the net angular momentum of the incoming flow is sufficiently small that flow circularizes inside the boundary condition. In these cases, we see narrow distributions with $|l|/ l_{\rm kep} \ll 1$. These cases exhibit higher accretion efficiencies in Figure \ref{fig:mdot}. In steeper-gradient cases, accretion is limited by angular momentum and we see overlapping, broad distributions of $|l|/ l_{\rm kep}$, with correspondingly low accretion efficiencies in Figure \ref{fig:mdot}. }
\label{fig:jdot}
\end{center}
\end{figure}

Turning now to the accretion of angular momentum, Figure \ref{fig:jdot} evaluates the distributions of the magnitude of specific angular momentum, $|l|$, of material absorbed by the sink boundary. These are normalized to the Keplerian specific angular momentum at the sink surface, $l_{\rm kep}$ \citep[for details see][section 4.3]{2015ApJ...803...41M}. In mild density gradient cases, accreted material has a relatively narrow distribution of specific angular momenta, with typical values much less than Keplerian. In these cases, the mass accretion rate is high (compare to Figure \ref{fig:mdot}) because the net angular momentum of the flow does not substantially oppose accretion when $|l|\ll l_{\rm kep}$. At higher values of the density gradient, the distributions of specific angular momenta of accreted material are much broader, with typical values of $|l|/ l_{\rm kep} \sim 0.5$. None of our simulations show signs of accreting material with nearly complete rotational support $|l| \sim l_{\rm kep}$, which makes sense because in a given timestep, any material that is fully rotationally supported will be unable to accrete. However, the $\gamma=4/3$ simulations show systematically higher specific angular momenta than those of the $\gamma=5/3$ simulations.

\subsection{Drag Forces}
With the inclusion of the gravitational restoring force of the primary star (and hydrostatic equilibrium pressure gradient) our simulations are much better suited to the evaluation of drag forces than those of \citet{2015ApJ...803...41M}.  
This section studies contributions to the force on an embedded object from gaseous dynamical friction and from the accretion of linear momentum by the sink boundary. The details of both terms are outlined in Section \ref{sec:diag}. 

We integrate the gravitational drag force (dynamical friction), given by equation \eqref{fdf}, over ten different volumes each timestep. The integration volumes are spherical shells with inner radius equal to $\Rs$ and outer radius evenly spaced in $\ln(r_{\rm out})$  between $0.1\Ra$ and $3.5\Ra$. The outer integration radii therefore are, $r_{\rm out}/\Ra \approx 0.1$,  0.15, 0.22,0.33, 0.49, 0.72, 1.06, 1.59, 2.36, 3.5. We find that in the steepest gradient cases, our results for the outer two radii are somewhat sensitive to the boundary position (whether the $y,z$ domain extends $\pm 3.5\Ra$ or $\pm 4\Ra$) because of the diode (no inflow) boundary conditions imposed. We therefore show results for integration radii only out to $1.59\Ra$ in what follows, which are converged with respect to boundary location in even the steepest gradient cases. 

The motivation for logarithmically spaced integration bins is that dynamical friction forces (in both gaseous and collisionless systems) in uniform media grow as $\propto \ln(r_{\rm out})$ \citep[e.g.][]{1943ApJ....97..255C,1999ApJ...513..252O}. By spacing our integration bins in this manner, each bin contains a similar contribution to the total force. The `advective' force due to the rate of accretion of linear momentum by the sink boundary is given by equation \eqref{pdotdiag} and is labeled $F_{ \dot p_x}$ here. In our coordinate system the dynamical friction force acts in the positive direction (and is therefore a drag), while net accretion generally acts in the negative direction (and is therefore a thrust) because most material is accreted from behind the embedded object.

\begin{figure}[tbp]
\begin{center}
\includegraphics[width=0.48\textwidth]{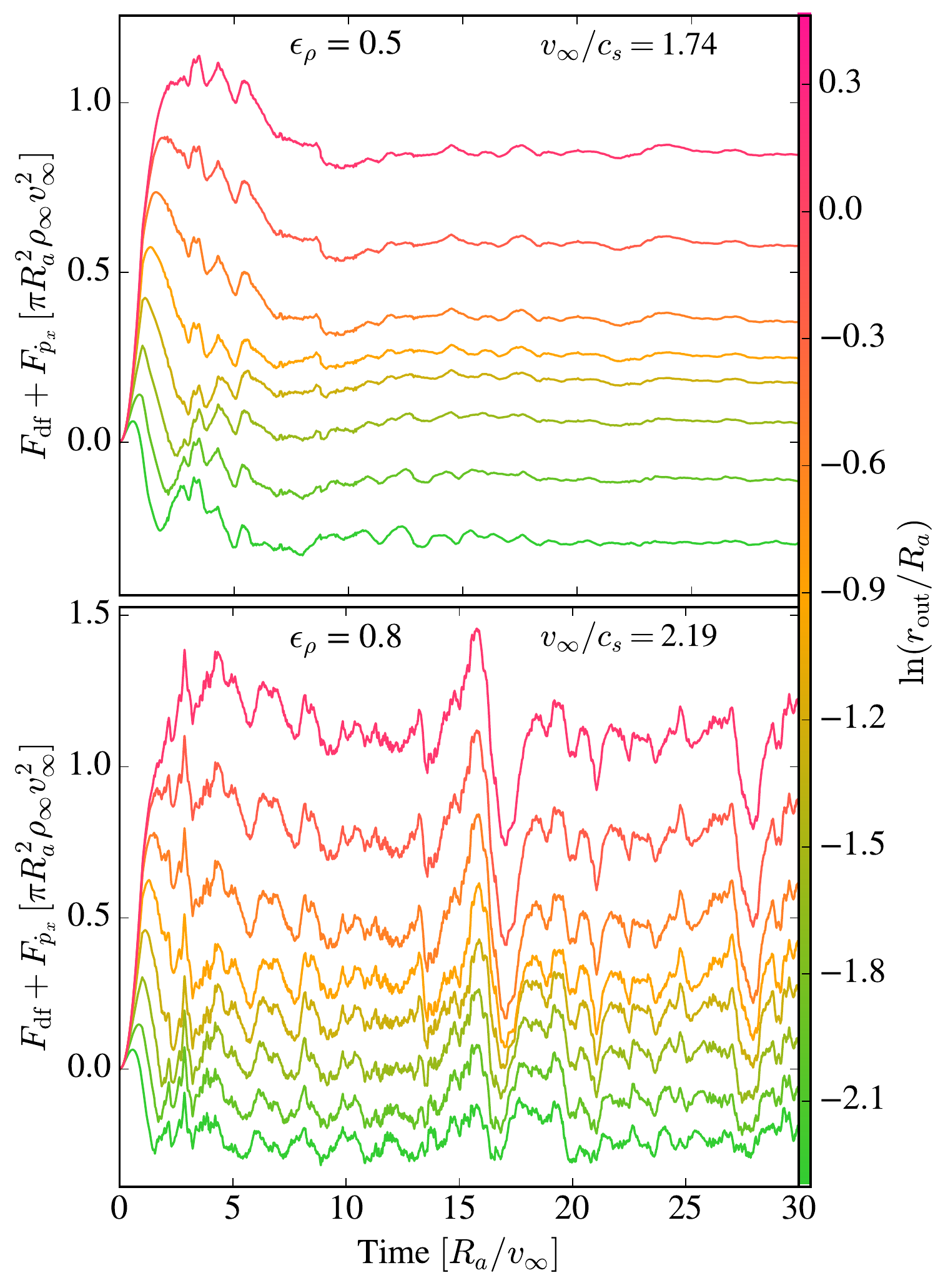}

\caption{Net drag forces (including contributions from dynamical friction and accretion of linear momentum) as a function of simulation time and dynamical friction integration radius (line color) for two example simulations. The initial flow is marked by rising drag forces as a wake sets up. The $\erho\approx0.5$ case shows relatively smooth drag force at late times, compared to the much more time-variable force of the $\erho\approx0.8$ simulation. This difference is reflected in the flow visualizations of Figures \ref{fig:g53z} and \ref{fig:g53y}, which show a transition from smooth to more turbulent post-shock flow as material transitions from circularizing inside to outside the sink boundary. In this measurement of the drag forces, we see variability imprinted on the net drag from small scales, particularly around $\sim 0.5 \Ra$, the standoff distance of the bow shock (orange line and below in the colorbar). }
\label{fig:drag_time}
\end{center}
\end{figure}

We begin by examining the net drag forces as a function of time in some example simulations. Figure \ref{fig:drag_time} plots the net force on the embedded object for each value of the outer integration radius of the dynamical friction force, $r_{\rm out}$, as denoted by line color. We show results for two simulations from the $\gamma=\gs=5/3$ simulation suite. These simulations have differing density gradients, $\erho\approx0.5$ and $\erho\approx0.8$. In both cases, we see an initial transient behavior while the flow sets up (the box crossing time is $\approx 8 \Ra/\vinf$), followed by a settling to a steady state. 

In the $\erho\approx0.5$ simulation, the flow visualizations in Figures \ref{fig:g53z} and  \ref{fig:g53y} show relatively smooth post-shock structure without substantial small-scale vorticity. As discussed in relation to Figure \ref{fig:jdot}, this is likely because the sink is large enough to swallow much of the circularizing material. As a result of the smooth wake, the dynamical friction force experienced by the embedded object is also relatively smooth. The net force is negative (a thrust) when the dynamical friction is only integrated to very small radii $\lesssim 0.3 \Ra$. When we include the contributions from progressively larger radii, $F_{\rm df}$ outweighs $F_{\dot p_x}$ and the net force is positive (a drag). 

The $\erho\approx 0.8$ panel of Figure \ref{fig:drag_time} shows many similarities to the $\erho\approx0.5$ panel, but exhibits substantially greater time variability. It is interesting to note that features in the variability overlap at many scales in the cumulative drag force plotted. Some variability is imprinted at the smallest scales, in particular, the short timescale (but relatively small amplitude) variation. The majority of the variability in the net drag is imposed at larger scales, of order the shock standoff distance. This occurs as vorticies shed in the wake cause some breathing and instability of the position of the bow shock. Since this large-scale flow instability is not present in the shallower-gradient cases, the drag force is much steadier. An interesting caveat to this point is that the size of the sink boundary likely plays a role in both the time variability of the drag force and which flow parameters generate highly variable post-shock regions. As shown in \citet[][Figure 7]{2015ApJ...803...41M}, smaller sink boundaries result in more vorticity in the post-shock region and more variable accretion. We can speculate that the boundary condition might imprint itself on the time variability of the drag force in a similar manner. 

The relative displacement of the lines in Figure \ref{fig:drag_time} is also informative.  To first order, the spacing in these simulations is relatively uniform, indicating that  the contribution to the dynamical friction drag is growing  approximately $\propto \ln (r_{\rm out})$. A transition to slightly larger spacings among the largest integration radii comes after passing the  approximate standoff distance of the bow shock -- in this simulation, $\sim 0.5\Ra$. Within the shock standoff radius, the density field is more symmetric (though still not entirely so) than integration radii that include the shock \citep[e.g.][]{2016A&A...589A..10T}.

\begin{figure}[tbp]
\begin{center}
\includegraphics[width=0.48\textwidth]{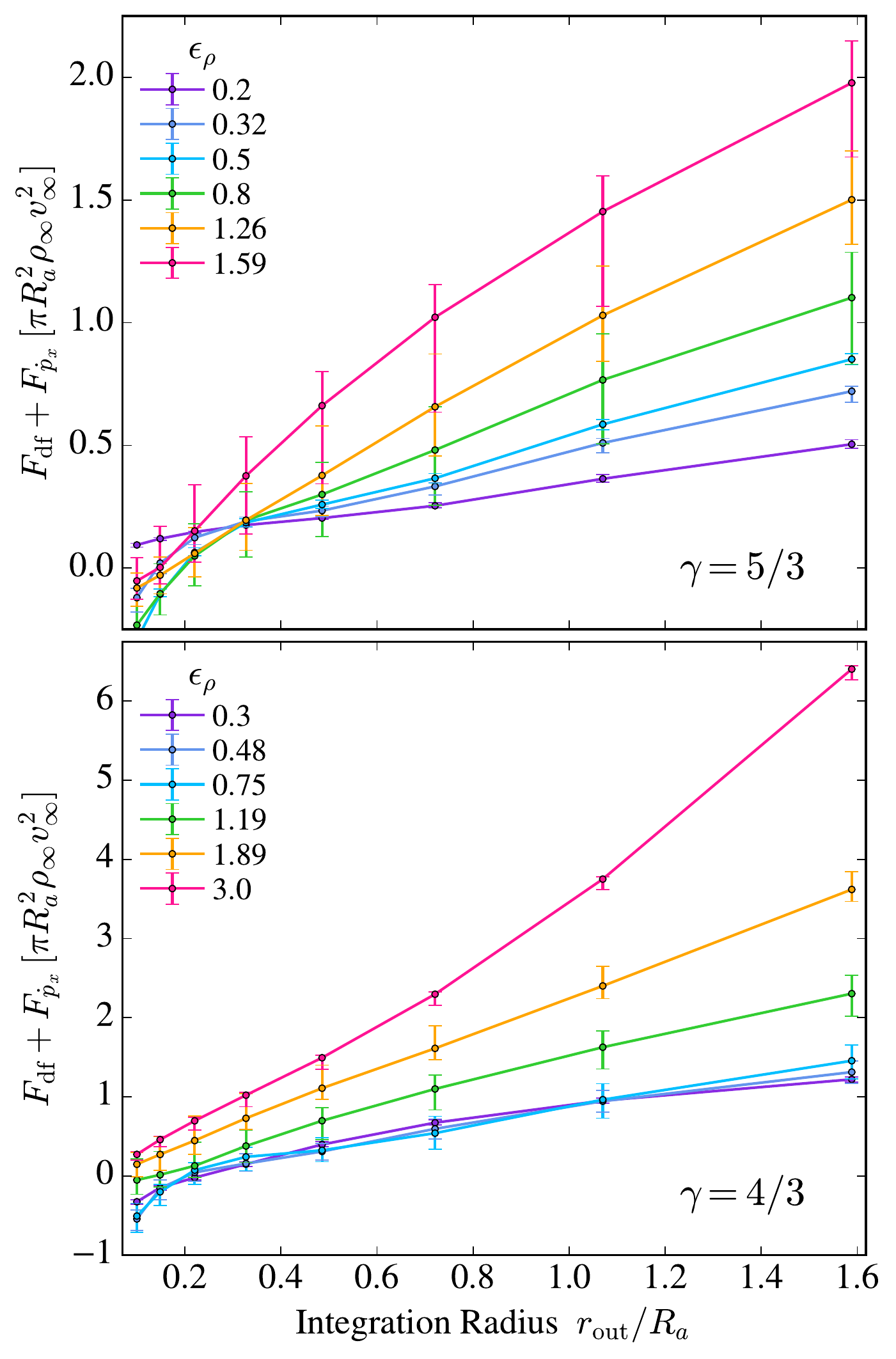}
\caption{Net drag forces (including dynamical friction and momentum accretion) for different dynamical friction outer integration radii and density gradient. Points and their errorbars show the median, along with fifth and ninety fifth percentile regions for times $10<t<30$ (after steady state is established) in our simulations.   }
\label{fig:drag_rad}
\end{center}
\end{figure}

Figure \ref{fig:drag_rad} shows the cumulative drag within different (logarithmically spaced) integration radii. As seen in the time series of Figure \ref{fig:drag_time}, the net drag is negative (a thrust) when the dynamical friction is only integrated out to a small radius. In the $\gamma=\gs=5/3$ case, we see that logarithmically spaced bins contribute roughly equally to the cumulative drag force. This implies that  the net drag is growing according to $\sim \ln \left( r_{\rm out} / r_{\rm in} \right)$,  as is the case for gas dynamical friction in homogenous media \citep{1999ApJ...513..252O,2016A&A...589A..10T}. The contribution of each increasing radius bin tells us something about the characteristic scale $r_{\rm in}$ in the dynamical friction force. A numerical comparison quickly reveals that the appropriate $r_{\rm in}$ is not $\Rs$, the radius of the inner boundary of our computational domain, but is instead something of order $\sim 0.5\Ra$.  \citet{2016A&A...589A..10T} find something qualitatively similar in their analysis, and they point out that this minimum scale is the standoff distance of the bow shock, because the density field becomes markedly more asymmetric outside this distance. This implies that our drag force results could be scaled to different maximum radii using a factor of $\sim \ln \left( r_{\rm out} / 0.5\Ra \right)$.

In the $\gamma=\gs=4/3$ simulation suite, the first characteristic we note is that the overall coefficients of drag are substantially larger than those in the $\gamma=\gs=5/3$ simulations. This is likely because the more compressible equation of state results in a higher density wake behind the embedded object, which then exerts a stronger gravitational deceleration on the embedded object's motion. The shallow gradient cases appear to grow roughly logarithmically, with $r_{\rm min}\sim 0.5 \Ra$ as in the $\gamma=\gs=5/3$ simulations. The steepest gradient cases of the $\gamma=\gs=4/3$ simulation suite show somewhat different behavior: the growth of $F_{\rm df}$ is superlogarithmic. 
\citet[][]{1999ApJ...513..252O}'s equation 13 shows that the logarithmic behavior comes, in part, from the constant opening angle of the Mach cone. The cases that grow more rapidly than $\ln(r_{\rm out})$ in our wind-tunnel calculation are those that show a flared wake due to the Mach number gradient discussed in Section \ref{sec:flow}. With a wider wake opening angle with increasing distance in these cases, the integrated dynamical friction drag grows faster than logarithmically in $r_{\rm out}$.

\begin{figure}[tbp]
\begin{center}
\includegraphics[width=0.48\textwidth]{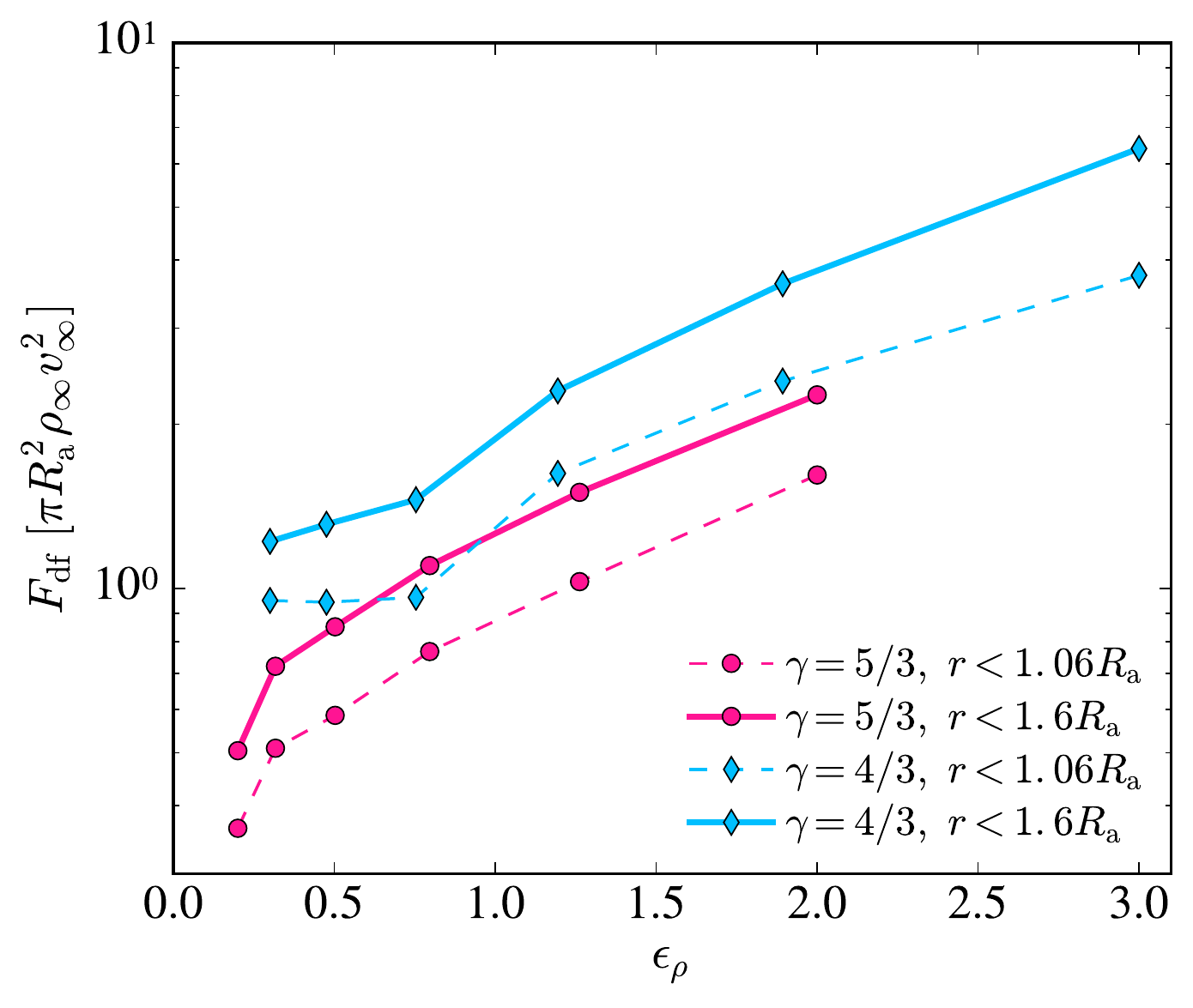}
\caption{Dynamical friction drag forces plotted versus density gradient for two integration radii, $1.06\Ra$ and $1.6\Ra$. The coefficient of drag is systematically higher in the more compressible $\gamma=4/3$ simulations because a higher density wake trails the embedded object. In all cases, the drag coefficient increases with density gradient, because dense material offset from the object's position in the $-y$-direction (toward the primary-star center) is focussed into the wake. }
\label{fig:drag_erho}
\end{center}
\end{figure}

Finally, we summarize our results for coefficients of dynamical friction as a function of density gradient and Mach number in Figure \ref{fig:drag_erho} (not including the advected momentum term). We see immediately that the coefficient of drag rises with increasing density gradient by a factor of several across the span of values we have simulated. Qualitatively, what we see is that in cases of steep density gradient, the drag force depends not only on the value of the density at the embedded object's position ($\rhoinf$), but also the sweep of higher densities within the accretion radius. The flow focusses this dense material into the wake of the embedded object, causing it to contribute to the net dynamical friction. We also can observe a sharp downturn in the drag coefficient in the shallowest gradient $\gamma=\gs =5/3$ case. This is readily explained by the low Mach number of this simulation, $\mach =1.1$, which allows pressure to partially resist the density asymmetry of the wake. \citet{1999ApJ...513..252O} discusses this effect extensively, and shows that the drag should behave $\propto \ln \left(1+\mach^{-2} \right)$ in the supersonic limit, thus decreasing steeply as $\mach\rightarrow1$ \citep[equation 15 in][]{1999ApJ...513..252O}.  The lowest Mach number simulation in our $\gamma=\gs=4/3$ suite has $\mach = 1.35$, so we do not expect (or see) as dramatic of a correction due to low flow Mach number.

We note here that the coefficients of drag and their dependence on $\erho$ derived here are different (though similar order of magnitude) from those in \citet[][Figure 13]{2015ApJ...803...41M}, both because of our updated formalism and different flow parameters (section \ref{sec:flowconditions}), and because of our corrected dynamical friction diagnostics described in section \ref{sec:diag}.  These updates represent a significant improvement in our ability to correctly asses the dynamical friction acting on the embedded object, and the difference of our new results reflects these changes.

\section{Implications for Common Envelope Inspiral}\label{sec:implications}

We have used idealized numerical simulations to study flow morphologies, as well as coefficients of drag and accretion for objects embedded in the common envelope. These quantities describe the transformation of an object and its orbit through the common envelope episode. Drag forces drive the orbital tightening, while flow convergence and mass accretion might transform the object itself.  

There are, of course, caveats associated with the simplifications we have made here. We have isolated particular flow conditions and measured steady-state rates of drag and accretion, but it is worth considering that steady state might not be realized during the complex and violent flow of a common envelope interaction. Among the potential concerns with extrapolating the results of these simulations is that the geometry of our simulations does not match that of the large-scale common envelope: we have adopted a cartesian geometry, where stars are spherical. We similarly disregard the effects of the rotating frame that co-moves with the embedded object. These simplifications almost certainly affect the exact numerical values derived for our coefficents of drag, particularly on scales $>\Ra$, which become similar to the binary separation, $a$, the scale where curvature becomes very important.  Similarly, by fixing the gas compressibility, $\gamma$, and studying two representative values of $4/3$ and $5/3$, we ignore  thermodynamic transitions that might result from the gas's passage through shocks and compression as it passes near the embedded object. 

Our coefficients of accretion have dependence on the size of the sink boundary, as documented in \citet{2015ApJ...803...41M}.  These rates should thus be treated as rates of flow convergence through a boundary of a particular size: if we are considering an embedded compact object, which might be orders of magnitude smaller, it is not obvious that all of the converging material will reach the embedded object's surface.  Secondly, not all objects are thermodynamically able to accrete from the common envelope gas. Accretion onto white dwarfs or main-sequence stars has no obvious cooling channel (photons will be trapped in the very dense flow) and therefore we probably should not expect mass accumulation on these objects despite flow convergence. On the other hand, for high enough accretion rates neutrinos can likely mediate the accretion luminosity of accretion onto neutron stars \citep{1991ApJ...376..234H,1993ApJ...411L..33C,1996ApJ...460..801F,2000ApJ...541..918B,2015ApJ...798L..19M}, and, lacking a surface, black holes will certainly accrete material passing through their horizons.

\begin{figure}[tbp]
\begin{center}
\includegraphics[width=0.48\textwidth]{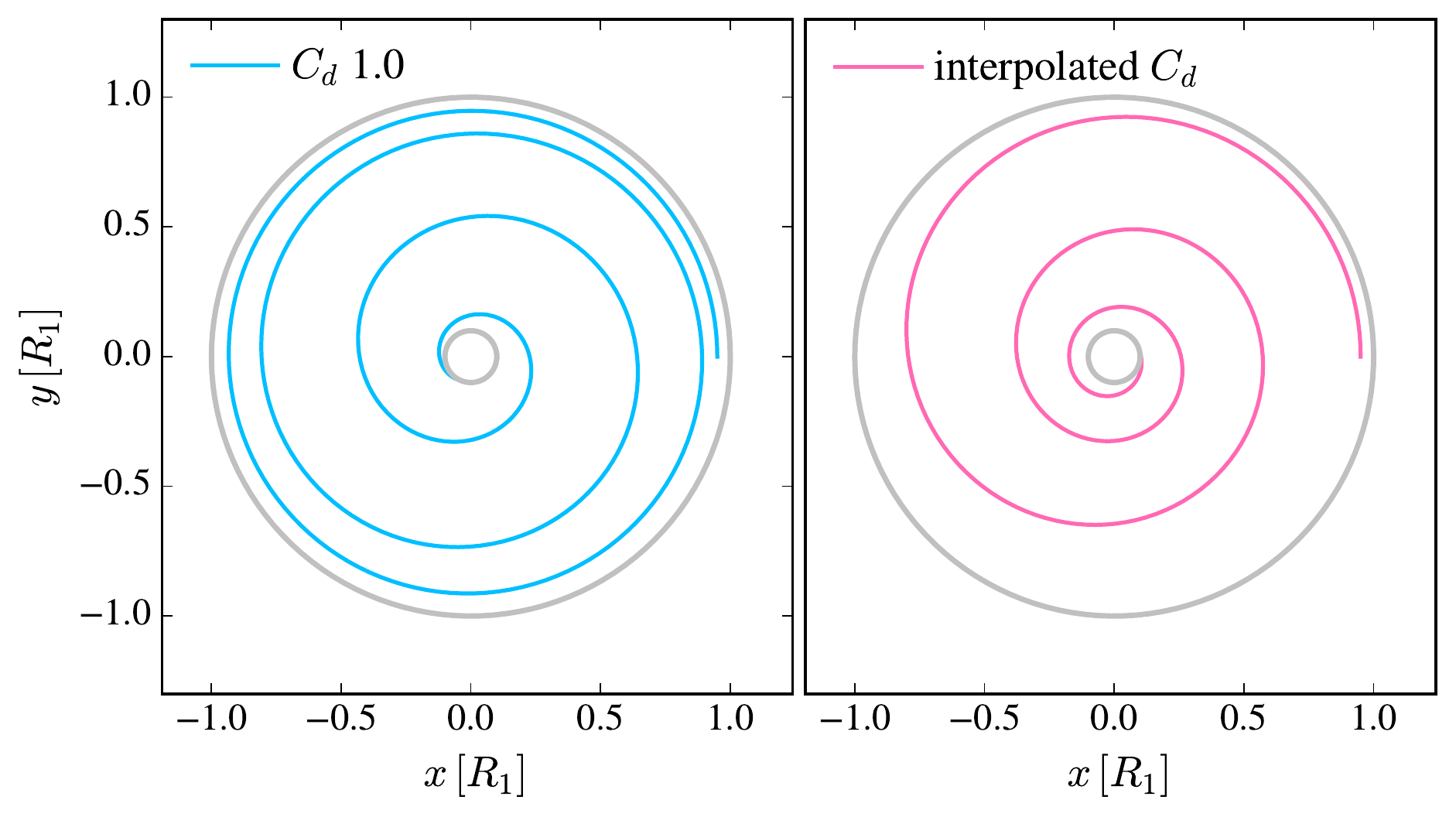}
\caption{Inspiral of a $0.3M_\odot$ secondary through a $3M_\odot$, $31R_\odot$ primary star's envelope. The two examples show a drag force applied with $F = C_{\rm d} \pi \Ra^2 \rho v^2$, where we adopt the Hoyle-Lyttleton value of $C_{\rm d}=1$ and a coefficient interpolated from our $\gamma=5/3$ simulation results of Figure \ref{fig:drag_erho}, for an integration radius of $1.6\Ra$.  With simulation coefficients applied, the initial orbital inspiral is much more rapid, while the late inspiral slows and wraps tighter than in the $C_{\rm d}=1$ case.  }
\label{fig:insp1}
\end{center}
\end{figure}

\begin{figure}[tbp]
\begin{center}
\includegraphics[width=0.48\textwidth]{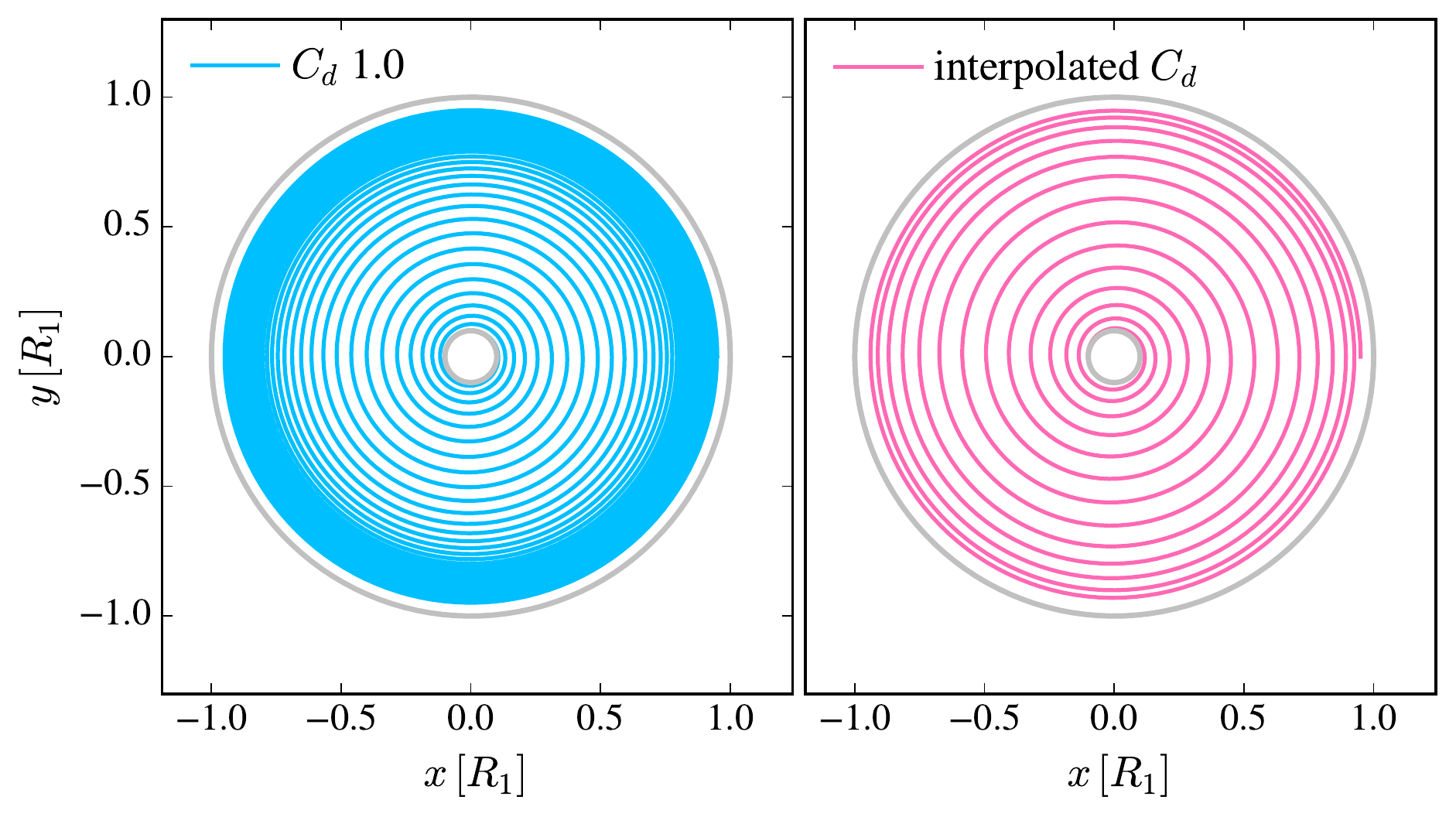}
\caption{Same as Figure \ref{fig:insp1} for an $8M_\odot$ secondary object and an $80M_\odot$, $720R_\odot$ primary star. Numerical coefficients of drag are interpolated from our $\gamma=4/3$ simulation suite in this calculation. }
\label{fig:insp2}
\end{center}
\end{figure}

Despite the remaining uncertainties, the coefficients of drag and accretion derived here carry lessons for the dynamics of common envelope episodes.  In \citet{2015ApJ...803...41M,2015ApJ...798L..19M}  we argued that the ratio of drag coefficient to accretion coefficient that arises from asymmetric flows in thecommon envelope implies that objects grow by at most a few percent during their inspiral. This qualitative conclusion remains unchanged despite our improved derivations of drag and accretion coefficients.

In Figures \ref{fig:insp1} and \ref{fig:insp2}, we illustrate the effect of including a coefficient of drag that varies with the flow parameters of the material that it is passing through. We use the primary-star profiles of Figure \ref{fig:stellarstructure}, and (as elsewhere in this paper), take a mass ratio $q=0.1$. We assume the primary star is initially non-rotating. We integrate the equation of motion of the secondary star relative to the enclosed mass of the primary, and add a drag force, $F_{\rm d} = C_{\rm d} \pi \Ra^2 \rho v^2$, where $C_{\rm d} = F_{\rm df} / \pi \Ra^2 \rhoinf \vinf^2$ is the coefficient of drag. We illustrate the influence of two choices: $C_{\rm d} = 1$, a Hoyle-Lyttleton drag force, and a numerically-derived $C_{\rm d}$ from our simulations (which comes from Figure \ref{fig:drag_erho}; here we take the force generated from $r<1.6\Ra$, and use the $\gamma = \Gamma_{\rm s} = 5/3$ case for the $3M_\odot$ primary and the $\gamma = \Gamma_{\rm s} = 4/3$ case for the $80M_\odot$ primary). Our example inspirals are initialized at $a=0.95 R_1$ and are integrated until $a=0.1R_1$. 

A priori, we might imagine that the initial inspiral of common envelope episodes is slow, taking many orbits while the secondary passes through the low-density atmosphere of the primary's envelope. Instead, with the realistic coefficients, the initial common envelope inspiral is substantially more rapid than with the Hoyle-Lyttleton force alone.  In the late inspiral, the drag force drops, and the orbits wrap tighter.  This result can be qualitatively understood in the context of our simulations: when an embedded object lies along a steep density gradient (where the scale height is small compared to $\Ra$), the object gravitationally focusses dense material from deeper in the stellar interior into its wake. This denser material (compared to the density at the secondary's position within the primary star) leads to a more massive wake, and a higher dynamical friction drag force. In terms of the flow streamlines shown in Figures \ref{fig:g53z} and \ref{fig:g43z}, the envelope gas contributing to the wake comes largely from dense material with impact parameters in the $-y$-direction in simulation coordinates -- toward the primary-star interior.  

One potential impact of the increased rapidity of early inspiral is on transients from the onset of a common envelope. The emergent class of luminous red novae transients has been associated with mass ejection in stellar merger and common envelope encounters \citep[see, e.g.][for recent examples]{2011A&A...528A.114T,2013Sci...339..433I,2015ApJ...805L..18W,2015A&A...578L..10K,2017ApJ...834..107B,2016MNRAS.458..950S,2016arXiv160501493M}. A rapid early inspiral would match the rapid lightcurve rise of some of these transients. For example, the M31 LRN 2015 outburst rose from detection to peak brightness in a timescale of order one binary orbital period. With a $\sim 3-5M_\odot$, $\sim 35R_\odot$ progenitor giant, this system had a primary star broadly similar to that shown in Figure \ref{fig:insp1} \citep{2015ApJ...805L..18W,2016arXiv160501493M}. This is a surprisingly rapid timescale when we compare to the slow early inspiral predicted by Hoyle-Lyttleton drag coefficients ($C_{\rm d} = 1$), but it is perhaps more consistent with our numerically derived coefficients, which show substantial inspiral in a single orbit. There remains much work to be done, though, to establish the mappings between orbit evolution, mass ejection, and light-curve generation in these events. 

The $q=0.1$ inspirals of Figure \ref{fig:insp1} and \ref{fig:insp2} differ qualitatively between the $3.0M_\odot$ primary and the $80M_\odot$ primary in the number of orbits elapsed during the inspiral. For the $3.0M_\odot$ primary, the secondary spirals to $a=0.1R_1$ in $\sim 4$ orbits, while in the  $80M_\odot$ case, the plunge takes $\sim 13$ orbits (with the interpolated drag coefficients). This difference reflects the difference in primary-star envelope structure. The density of the $80 M_\odot$ red supergiant envelope is very low, because radiation pressure (and the fact that the star is nearly at the Eddington limit) inflates the envelope \citep[e.g.][]{2017A&A...597A..71S}. One consequence of this difference might be that the embedded star orbits through material that it has disturbed (or shock heated) in previous passages if the change in separation between orbits is not greater than the typical bow shock scale. In other words, when $\dot a P_{\rm orb} \lesssim \Ra$, we can expect that the envelope is disturbed from its initial state on subsequent orbits.\footnote{see, \citet{1988ApJ...329..764L} and \citet{1993PASP..105.1373I} for similar considerations based on stellar envelope profiles. } In these cases, we might expect some departure from our common envelope wind tunnel flow relations. The exact extent to which this effect is important will depend on the spherical geometry of the flow \citep[e.g.][]{2007JKAS...40..179K,2007ApJ...665..432K,2008ApJ...679L..33K,2010ApJ...725.1069K,2011ApJ...739..102K} and is difficult to asses within the context of the simulations presented here.

\section{Conclusions}\label{sec:conclusion}

In this paper, we have shown characteristic relationships between the density scale height and Mach number in the common envelope based on the primary star's structure, and we have studied three-dimensional realizations of these gas flows in an idealized ``wind tunnel'' setup. We draw several key conclusions from this work:

\begin{enumerate}
\item We have derived relationships for dimensionless flow scales that generically characterize common envelope flows. In particular, equations \eqref{mach} and \eqref{erho} relate flow Mach numbers and density gradients in terms of binary mass ratio, envelope structure, and relative velocity. 

\item These relationships between flow parameters affect common envelope flow morphologies in a correlated way. Low Mach number flows tend to have mild density gradients, while high Mach number flows are always accompanied by steep gradients. 

\item Density gradients in common envelope flows limit mass accretion toward the embedded objects to a fraction of $\mhl = \pi \Ra^2 \rhoinf \vinf$, where density and velocity are defined at the location of the embedded object within the envelope.

\item Dynamical friction (gravitational) drag forces are enhanced by steep density gradients compared to the Hoyle-Lyttleton drag force, $\mhl \vinf = \pi \Ra^2 \rhoinf \vinf^2$, because of  the contribution from dense material $(\rho\gg\rhoinf)$ that is  focussed into the wake of the embedded object from deeper within the stellar interior (the $-y$-direction in our wind tunnel setup).  These conditions are particularly relevant near stellar envelope limbs, implying more rapid orbital evolution at the onset of common envelope interactions than predicted from the Hoyle-Lyttleton force alone, with potential implications for the timescale of associated transients.

\end{enumerate}

There remain many future questions to address, even in the context of simplified studies of flow within a common envelope ``wind tunnel.''  In future simulations, we imagine it will be particularly worthwhile to consider flow properties in cases of partial synchronization between the primary-star envelope and the secondary's orbital motion ($f_{\rm k} <1$),  the role of equation of state, and of non-accreting secondary stars. To understand open questions about the transition from dynamical plunge to subsonic motion and stabilized inspiral \citep{2001ASPC..229..239P,2013A&ARv..21...59I,2016MNRAS.462..362I,2016MNRAS.461..486K,2017MNRAS.464.4028I}, it is likely critical to move beyond the wind tunnel formalism established here to capture the details of the passage of objects through envelope gas, which they have already perturbed. However, even shock-heated material will retain the relationships between density gradient and flow Mach number described in Section \ref{sec:flowconditions} if it is in (approximate) hydrostatic equilibrium.

\begin{acknowledgements}
We acknowledge helpful discussions with G. Laughlin, D. Lee, D. Lin, E. Ostriker, M. Rees, J. Samsing, and J. Stone, which helped shape and improve this work. 
We also acknowledge helpful feedback from anonymous referees. 
The software used in this work was in part
developed by the DOE-supported ASCI/Alliance Center for
Astrophysical Thermonuclear Flashes at the University of
Chicago. Simulation visualizations and analysis were made
possible using the {\tt yt} toolkit \citep{2011ApJS..192....9T}.
This research made use of {\tt astropy}, a community developed
core Python package for Astronomy \citep{2013A&A...558A..33A}. 
The calculations for this research were carried out in part on the UCSC supercomputer Hyades, which is supported by the National Science Foundation (award number AST-1229745) and UCSC. 
MM is grateful for support for this work provided by NASA through Einstein Postdoctoral Fellowship grant number PF6-170155 awarded by the Chandra X-ray Center, which is operated by the Smithsonian Astrophysical Observatory for NASA under contract NAS8-03060.
AA gratefully acknowledges support from the NSF REU program LAMAT at UCSC, a UCSC Undergraduate Research in the Sciences Award, and the California Space Grant Consortium (CaSGC) Undergraduate Research Opportunity Program.
AMB acknowledges UCMEXUS-CONACYT Doctoral Fellowship.
PM is supported by an NSF Graduate Research Fellowship and a Eugene Cota-Robles Graduate Fellowship. 
ER-R acknowledges financial support from the Packard Foundation, Radcliffe Institute for Advanced Study and NASA ATP grant NNX14AH37G. 
\end{acknowledgements}

\bibliographystyle{aasjournal}
\bibliography{ce}

\end{document}